\documentclass[11pt]{article}
\usepackage[margin=1in]{geometry}
\usepackage{amsmath,amssymb}
\usepackage{graphicx}
\usepackage{booktabs}
\usepackage{natbib}
\usepackage{hyperref}
\usepackage{longtable}
\usepackage{caption}
\usepackage{subcaption}
\usepackage{xcolor}
\usepackage[normalem]{ulem} 
\usepackage{array}
\usepackage[useregional]{datetime2}
\usepackage{authblk}

\title{Feeling Left Behind? Territorial Disadvantage, Well-Being and Cohesion across European Regions}

\author[1]{Stefano M.\ Iacus}
\author[2]{Giuseppe Porro}
\author[3,4]{Haodong Qi}
\author[5]{Devika Jain}

\affil[1]{Institute for Quantitative Social Science, Harvard University, Cambridge, MA 02138, USA}
\affil[2]{Department of Law, Economics and Culture, University of Insubria, Como, Italy}
\affil[3]{Stockholm University Demography Unit, Stockholm, Sweden}
\affil[4]{Malm\"o University, Department of Global Political Studies, Malm\"o, Sweden}
\affil[5]{Center for Geographic Analysis, Harvard University, Cambridge, MA 02138, USA}

\date{}

\begin{document}
\maketitle

\begin{abstract}
Left-behind places are usually identified using economic, demographic and accessibility indicators, but these may not align with well-being and social cohesion. We combine ten years of territorial indicators with subjective measures derived from georeferenced social-media data for over 1,200 NUTS-3 regions in 28 European countries (2013–2023). Regional profiles and within-between panel models reveal uneven relationships: disadvantaged regions can display contrasting levels of well-being and cohesion, national context alters cross-country associations, and between-region differences diverge from within-region change. The findings show that aggregate measures conceal distinct forms and trajectories of territorial disadvantage.

\end{abstract}
\noindent\textbf{Keywords:} left-behind places; territorial disadvantage; regional inequality; subjective well-being; social cohesion; NUTS-3 regions

\clearpage

\section{Introduction}

Growing spatial inequalities have returned territorial disadvantage to the centre of academic and policy debate. Over the last decade, the notion of \emph{left-behind places} has been applied to former industrial areas, shrinking rural regions and peripheral territories affected by combinations of economic stagnation, deindustrialisation, demographic decline, ageing, outmigration, poverty, weak accessibility and deteriorating services \citep{RodriguezPose2018,MacKinnonEtAl2022,FiorentinoEtAl2024,PikeEtAl2024}. Yet left-behindness is not reducible to low income or a simple core--periphery divide. It is a relative, multidimensional and dynamic condition whose dimensions and trajectories vary across places \citep{ComimEtAl2024,Bernard2025,PeranchoEtAl2025}. Using economic, demographic, social and accessibility indicators, \citet{velthuis2025} identify three varieties among EU15 NUTS-3 regions: economic decline and deindustrialisation, demographic decline and ageing, and disconnection and high poverty. Their typology shifts attention from identifying poorly performing regions to understanding how different forms of disadvantage accumulate and persist.

The objective geography of left-behindness does not, however, reveal how residents experience place. Recent research has begun to relate multidimensional disadvantage to subjective feelings and attitudes \citep{PanoriEtAl2025}, but has focused mainly on political discontent, institutional distrust and support for populist or anti-establishment movements \citep{RodriguezPose2018,MacKinnonEtAl2024,Bernard2025}. Much less is known about subjective well-being, generalized social trust and belonging. These outcomes are conceptually distinct: distrust in institutions need not imply distrust in other people, and dissatisfaction with government need not imply weak community attachment.

Competing mechanisms may connect territorial conditions to these outcomes. Economic stagnation, declining opportunities, poverty, outmigration and weak accessibility can reduce well-being and social trust, while prolonged neglect may alter residents' relations with institutions and their wider social environment \citep{MacKinnonEtAl2022,MacKinnonEtAl2024}. Conversely, disadvantage can coexist with strong attachment and solidarity. In shrinking rural regions, familiarity, continuity and established social relations may sustain attachment among those who remain \citep{VanDerStarHochstenbach2022}; local networks, anchor institutions and traditions of collective action may likewise support community resilience \citep{McAreavey2022,HalfordEtAl2025}. These resources should not be romanticised, but they imply that material disadvantage, well-being, ill-being, trust and belonging need not share the same geography or be collapsed into a single measure.

We investigate these relationships by combining a longitudinal geography of objective disadvantage with geographically disaggregated subjective indicators. We extend the multidimensional approach of \citet{velthuis2025} from the EU15 to 28 European countries and from a cross-section to a 2013--2023 NUTS-3 panel. We retain distinct regional varieties and construct a continuous left-behindness score from economic, demographic, social and accessibility indicators. Subjective outcomes come from the Human Flourishing Geographic Index, derived from georeferenced Twitter/X discourse, and cover fourteen dimensions of personal well-being, social cohesion and ill-being \citep{HFGI,iacus2026}. They capture geographically structured \emph{expressed} states and attitudes among observed users, not representative population estimates or direct measures of perceived left-behindness.

The panel allows us to separate relationships that cross-sectional analysis conflates. Regions in different countries vary in institutions, economic structures, language, culture and patterns of expression, while persistent differences between regions need not correspond to changes within the same region over time. We therefore distinguish cross-national comparisons, differences between regions within countries and temporal changes within regions.

The analysis asks four questions. First, how closely does objective territorial disadvantage correspond to well-being, ill-being and social cohesion? Second, how do these relationships change across cross-national, within-country and within-region comparisons? Third, how are temporal changes in left-behindness and human flourishing associated? Fourth, do economic decline and deindustrialisation, demographic decline and ageing, and disconnection and poverty have different subjective correlates? We also test whether belonging and cohesion vary nonlinearly with disadvantage.

The paper makes three contributions. It extends work linking objective left-behindness to subjective feelings \citep{PanoriEtAl2025} by considering well-being, ill-being and social cohesion. It introduces broader geographical coverage and a ten-year NUTS-3 panel relative to \citet{velthuis2025}, enabling separation of cross-national, within-country and within-region relationships. Finally, it combines conventional territorial indicators with fine-grained measures derived from geographically situated public discourse rather than predefined survey questions. Although these measures complement rather than replace representative surveys, relating them to aggregate left-behindness, its components and regional varieties reveals how different forms and trajectories of disadvantage correspond to distinct subjective configurations.
\section{Data and Methods}
\label{sec:data}

We combine two sets of data to study the relationship between objective territorial disadvantage and subjective expression. On the objective side, we reconstruct as closely as possible the indicators used by \cite{velthuis2025}, while extending their geographical and temporal coverage to 28 European countries and the period 2013--2023. The analysis covers 24 EU member states, together with the United Kingdom, Iceland, Norway and Switzerland. Cyprus, Malta and Luxembourg are excluded because HFGI indicators are not available for these countries in the dataset used in this study. The resulting harmonised regional panel is itself one contribution of this study.
On the subjective side, we use Twitter/X-based indicators from the Human Flourishing Geographic Index (HFGI) dataset \citep{HFGI}, derived from the analysis of a few billion georeferenced tweets from the \citet{geotweet2016} repository.

We align all this data to the 2021 NUTS-3 classification \citep{eurostat_nuts2021}, the finest EU-wide administrative geography available to us.

\subsection{Objective measurements of left-behindness}
\label{sec:data-objective}

Three sources cover the ten variables \cite{velthuis2025} used to construct their typology (Table~\ref{tab:variable-sources}):
ARDECO, the European Commission's Annual Regional Database
\citep{ardeco}, for GDP per capita, employment, population and industrial-employment
share, at NUTS-3 back to 1991; Eurostat, directly, for net migration and
old-age dependency \citep{eurostat}; and ESPON, the European Observation
Network for Territorial Development and Cohesion \citep{espon}, for
 poverty and accessibility. \cite{velthuis2025} also use
a NUTS-3 travel-time-to-shops measure and a direct youth-migration rate. As the level of data granularity is below what is needed for our analysis, we
substitute a related but distinct ESPON accessibility index for the former (no direct travel-time equivalent exists in ESPON), and construct youth
migration ourselves as explained below, since Eurostat does not publish the single-year-
of-age population at NUTS-3 level.

\begin{table}[htbp]
\centering
\caption{Objective variables, construction and source}
\label{tab:variable-sources}
\small
\begin{tabular}{lll}
\toprule
Variable & Construction & Source \\
\midrule
GDP per capita relative to national mean & ratio to national GDP/head, same year & ARDECO \\
GDP per capita growth differential & region's growth $-$ national growth, since 1991 & ARDECO \\
Employment growth differential & region's growth $-$ national growth, since 1991 & ARDECO \\
Population growth differential & region's growth $-$ national growth, since 1991 & ARDECO \\
Industrial employment share change & region's own point change since 1995 & ARDECO \\
Net migration rate & direct & Eurostat \\
Old-age dependency ratio & direct, single year & Eurostat \\
Youth net migration rate & cohort-component estimate, 2014--2019 & Eurostat (derived) \\
At-risk-of-poverty rate & recovered/imputed series & ESPON \\
Accessibility index & recovered series & ESPON \\
\bottomrule
\end{tabular}
\end{table}

\paragraph{Indicator construction.}
To ensure comparability with \citet{velthuis2025}, we reproduce each indicator in the same functional form. Regional GDP per capita relative to the national mean is expressed as a ratio; GDP per capita, employment and population growth as regional--national growth differentials; industrial-employment change as a regional point change; and migration, old-age dependency, poverty and accessibility as levels. Long-run growth is measured from 1991, or the earliest available national year, while industrial-employment change starts in 1995. Because Eurostat provides only five-year age bands at NUTS-3 level, youth migration is estimated using a five-year cohort-component method.

\paragraph{Data harmonisation and controls.}
ESPON poverty, income and accessibility indicators are harmonised from their original NUTS vintages to the NUTS~2021 geography using GISCO boundary crosswalks \citep{gisco}. Remaining poverty gaps are estimated with a validated a spatial generalized additive model (GAM) \citep{wood2003thinplate,wood2017gam}, while observed values are retained wherever available. The statistical models also include Eurostat's time-invariant urban--rural classification \citep{gisco,giscoR}. Full definitions, formulas, coverage diagnostics, imputation procedures and validation results are reported in Supplementary Material Section~\ref{sec:supp-indicator-construction}.

\subsection{Measures of subjective well-being and social cohesion}
\label{sec:data-subjective}
The social-media-based indicators used here are drawn from the broader set developed within the \textit{Human Flourishing Geographic Index} (HFGI) project.
The Human Flourishing Geographic Index dataset \citep{HFGI} is conceptually inspired by Harvard’s Human Flourishing Program \citep{VanderWeele2017}, which defines flourishing as a multidimensional construct across six domains: happiness and life satisfaction, mental and physical health, meaning and purpose, character and virtue, close social relationships, and material and financial stability. Whereas the Global Flourishing Study provides cross-country, wave-based (starting from 2023, essentially when our analysis ends) measures from survey data, the HFGI dataset offers high-resolution indicators at GADM administrative boundaries (version 2.8) \citep{gadm} and monthly and yearly frequency for the period 2013-2023 (June). As mentioned, it has been obtained  analyzing a few  billion of geo-referenced tweets from the Harvard CGA Geotweet Archive \citep{geotweet2016}, classified using a  fine-tuned large language model (LLM) \citep{finetuning2026}, to generate indicators corresponding to the Global Flourishing Study framework. 
The  content of the HFGI dataset and how it has been derived and validated, is illustrated in \cite{iacus2026}. Section~\ref{sec:issues} os the Supplementary Material reminds of limitations and potential biases in social media derived indicators.
Of the 46 HFGI dimensions, we selected fourteen covering both well-being/cohesion and
ill-being. These are: \emph{happiness, life satisfaction, hope, optimism, resilience}, for well-being; \emph{belonging, government
approval,  trust, trusted} for cohesion and belonging, and 
\emph{anxiety, depression, discrimination, loneliness, fear of the future} for ill-being. Every
dimension shares an identical structure in time and space, and each varies on the interval $[-1,+1]$, with 0 denoting a neutral value. The indicators are mapped and aggregated to NUTS-3 as described in detail in Section~\ref{sec:crosswalk} of the Supplementary Material.

\paragraph{Composite indices.} Each of the 14 raw dimensions is $z$-scored
across region-years, $\tilde y^{(k)}_{it} = (y^{(k)}_{it} - \bar
y^{(k)})/s_{y^{(k)}}$. Four composites are built from these standardized
components. The nine well-being dimensions are further split into two
constructs that not necessarily move the same way: \emph{Cohesion}, the four
relational/institutional dimensions (belonging, government approval, trust,
trusted), and \emph{Wellbeing}, the five personal/psychological dimensions
(happiness, life satisfaction, hope, optimism, resilience).

\begin{align}
\text{Composite}_{it} &= \frac{1}{14}\left(
  \sum_{k \in \text{well-being (9)}} \tilde y^{(k)}_{it}
  \;-\; \sum_{k \in \text{ill-being (5)}} \tilde y^{(k)}_{it}
\right) \\
\text{Cohesion}_{it} &= \frac{1}{4}\sum_{k \in \text{relational (4)}} \tilde y^{(k)}_{it} \\
\text{Wellbeing}_{it} &= \frac{1}{5}\sum_{k \in \text{personal (5)}} \tilde y^{(k)}_{it} \\
\text{Illbeing}_{it} &= \frac{1}{5}\sum_{k \in \text{ill-being (5)}} \tilde y^{(k)}_{it}
\end{align}

The pooled Composite reverses the five ill-being dimensions before averaging,
so that all 14 point in the same direction (higher = better) before pooling.
Cohesion, Wellbeing and Illbeing are each kept in their \emph{own} group's
natural polarity. Illbeing is \emph{not} reversed, so a higher value still
means \emph{more} ill-being on its own scale. The purpose of reporting these
three group-level composites alongside the pooled one is to make any
offsetting relationship between the underlying groups directly visible. If
objective left-behindness were positively associated with the well-being
side but this were being cancelled out, in the pooled Composite, by an
equally-sized positive association with the ill-being side, only the group
composites, not the pooled one, would reveal it. Splitting Cohesion from
Wellbeing serves the same purpose one level down: the relational and the
personal-psychological dimensions are conceptually distinct (shared social
attachment vs. an individual's own state) and, as Section~\ref{sec:results-1a}
shows, do not always move together, something a single nine-dimension
average would hide.

\subsection{Objective typologies and subjective profiles}
\label{sec:cluster}
This section examines how the objective typology proposed by \cite{velthuis2025} relates to the subjective profiles observed in the HFGI indicators.
Through cluster analysis, \cite{velthuis2025} built a 6-group typology of countries, with 3 types classified as left-behind (\emph{Economic decline and deindustrialisation}, \emph{Demographic decline and ageing}, and \emph{Disconnected, high poverty}) and 3 classified as not left-behind (\emph{Long-term economic prosperity}, \emph{High growth}, and \emph{Relative economic and demographic stability}).
We first assess how closely we can reproduce the reference typology, while recognising that our inputs are not perfectly aligned with those of the original study: our data cover more countries, span a longer period, and require the substitutions and estimations described above.

We proceed as in the original paper. Let $X$ be the $N \times 10$ matrix of the objective variables in
Table~\ref{tab:variable-sources}. Columns are standardized, $z_{ij} = (X_{ij} - \bar
X_j)/s_j$.

Starting with the 15 countries considered by the authors, a $k$-means
clustering method is applied with $k=6$, using the same point-in-time
(2018) and dynamic (1991--2018, 2014--2019) variable windows as
\citep{velthuis2025}. Multivariate outliers are removed via Mahalanobis
distance, as in the reference work (36 of 1,093 regions removed here).
On this clean EU-15 benchmark, reconstructed labels agree with
\citep{velthuis2025}'s own per-region labels for 47.3\% of the 1,033
matched regions. Collapsing to a binary left-behind/not-left-behind
classification, agreement is 68.8\% (Cohen's $\kappa = 0.387$).
These figures are acceptable but not especially strong, we therefore move to a $k=8$ solution for the whole set of 28-countries and temporal range.

Redoing the exercise with $k=8$, we reconstruct all six types proposed by
\cite{velthuis2025} and identify two additional clusters with no
 analogue (a non-left-behind ``Hyper-growth'' variety and a
two-region GDP outlier). In this case, full-label agreement rises to
54.3\%. Collapsing to a binary left-behind/not-left-behind classification,
agreement also rises to 76.1\% (Cohen's $\kappa = 0.524$). Restricting to the 396 regions both typologies flag as left-behind,
agreement on \emph{which} left-behind variety each region is goes up to 80.8\%. Therefore, this tiny relaxation of the clustering classification from 6 to 8 clusters allows for a more coherent description of the concept of left-behindness.

\subsection{A left-behindness score}

We complement the discrete typology with a continuous left-behindness score based on principal component analysis (PCA) \citep{jolliffe2002pca}. Let $Z$ be the standardized indicator matrix and $\mathbf{z}_i^\top$ its row for region $i$. The raw score is the projection onto the leading eigenvector $\mathbf{w}_1$ of $\mathrm{cov}(Z)$: $s_i = \mathbf{z}_i^\top \mathbf{w}_1$.
Because the sign of a principal component is arbitrary, we orient the score so that higher values indicate greater left-behindness, using poverty as the unambiguous ``higher = more left-behind'' anchor:
$\tilde{s}_i = \mathrm{sign}\big(\mathrm{cor}(S,\ \mathrm{poverty})\big)
\cdot s_i$, with
$S = (s_1, \dots, s_N)$.
This changes only the score's orientation; the PCA-determined variable weights remain unchanged. The oriented score is then standardized:
\begin{equation}
\mathrm{LB}_i = \frac{\tilde{s}_i - \bar{\tilde{s}}}{\sigma_{\tilde{s}}}.
\label{eq:lb-score}
\end{equation}
For the panel models in Section~\ref{sec:model}, we construct a dynamic version by recomputing the growth differentials annually while holding $\mathbf{w}_1$ fixed, ensuring comparability over time.
Table~\ref{tab:cluster-profiles} reports the mean LB score for each of the eight clusters and, where possible, its match to the six types identified by \citet{velthuis2025}.

\begin{table}[htbp]
\centering
\caption{Final objective typology, $k=8$, all 28 countries, ordered by left-behindness score}
\label{tab:cluster-profiles}
\small
\resizebox{\textwidth}{!}{%
\begin{tabular}{lrrcl}
\toprule
Cluster & $n$ & LB score & Left-behind? & Matched Velthuis et al.\ type \\
\midrule
Disconnected, high poverty & 124 & $+1.42$ & Yes & Disconnected, high poverty \\
Demographic decline and ageing & 232 & $+0.94$ & Yes & Demographic decline and ageing \\
Relative economic and demographic stability & 194 & $+0.54$ & No & Relative economic and demographic stability \\
Economic decline and deindustrialisation & 344 & $-0.15$ & Yes & Economic decline and deindustrialisation \\
High growth & 336 & $-0.66$ & No & High growth \\
Long-term economic prosperity & 119 & $-1.39$ & No & Long-term economic prosperity \\
Hyper-growth (Western Europe) & 20 & $-2.31$ & No & -- (no match) \\
Extreme outlier & 2 & $-5.79$ & No & -- (no match; per capita GDP anomaly) \\
\bottomrule
\end{tabular}%
}
\end{table}

Cluster labels and the continuous LB score provide complementary classifications. In particular, ``Economic decline and deindustrialisation'' is categorically left-behind despite its slightly negative mean score. The categorical label reflects the cluster's multidimensional profile, whereas LB is a single linear projection dominated by poverty and old-age dependency. Industrial decline loads more weakly and, in the cross-national distribution, in the opposite direction.

Table~\ref{tab:table2} compares our 2018 EU15 cluster means with the published means for the six matched types in \citet{velthuis2025}.

\begin{table}[htbp]
\centering
\caption{Per-cluster means vs.\ Velthuis et al.'s (2025) own published Table 2 means, 2018 EU-15 matched clusters (ours / theirs)}
\label{tab:table2}
\scriptsize
\resizebox{\textwidth}{!}{%
\begin{tabular}{lrrrrrrrr}
\toprule
Cluster & $n$ (ours/theirs) & pc-GDP rel. & pc-GDP gr.\ (pp) & Emp.\ gr.\ (pp) & Ind.\ chg (pp) & Net migr.\ (\textperthousand) & Old-age dep.\ (\%) & Poverty (\%) \\
\midrule
Demographic decline and ageing & 259/164 & 0.80/0.75 & $-5.5$/$-3.6$ & $-13.8$/$-26.1$ & $-4.2$/$-7.3$ & 2.0/17.2 & 39.7/41.8 & 15.0/17.2 \\
Disconnected, high poverty & 80/101 & 0.73/0.73 & $-7.1$/$-3.3$ & $-1.5$/$-11.7$ & $-2.5$/$-2.5$ & $-2.0$/$-6.0$ & 30.4/36.7 & 28.7/28.2 \\
Economic decl.\ and deindustr.\ & 309/232 & 0.88/0.84 & $-16.4$/$-13.6$ & $-4.7$/$-7.6$ & $-8.2$/$-12.2$ & 3.7/32.2 & 31.4/30.3 & 12.9/19.6 \\
High growth & 253/188 & 0.87/1.05 & $-1.9$/24.6 & 15.4/21.9 & $-3.8$/$-5.5$ & 7.2/67.1 & 32.3/30.1 & 12.3/13.0 \\
Long-term econ.\ prosperity & 94/85 & 1.50/1.50 & $-0.8$/$-2.1$ & 8.7/6.5 & $-5.5$/$-9.0$ & 8.2/72.1 & 27.4/25.1 & 12.0/18.5 \\
Relative econ./demogr.\ stability & 62/313 & 0.68/0.80 & $\mathbf{+204.4}$/$\mathbf{-5.0}$ & $-35.3$/0.4 & 1.1/$-5.5$ & 3.4/46.2 & 43.0/35.5 & 17.1/13.8 \\
\bottomrule
\end{tabular}%
}
\end{table}

\begin{figure}[htbp]
\centering
\includegraphics[width=\textwidth]{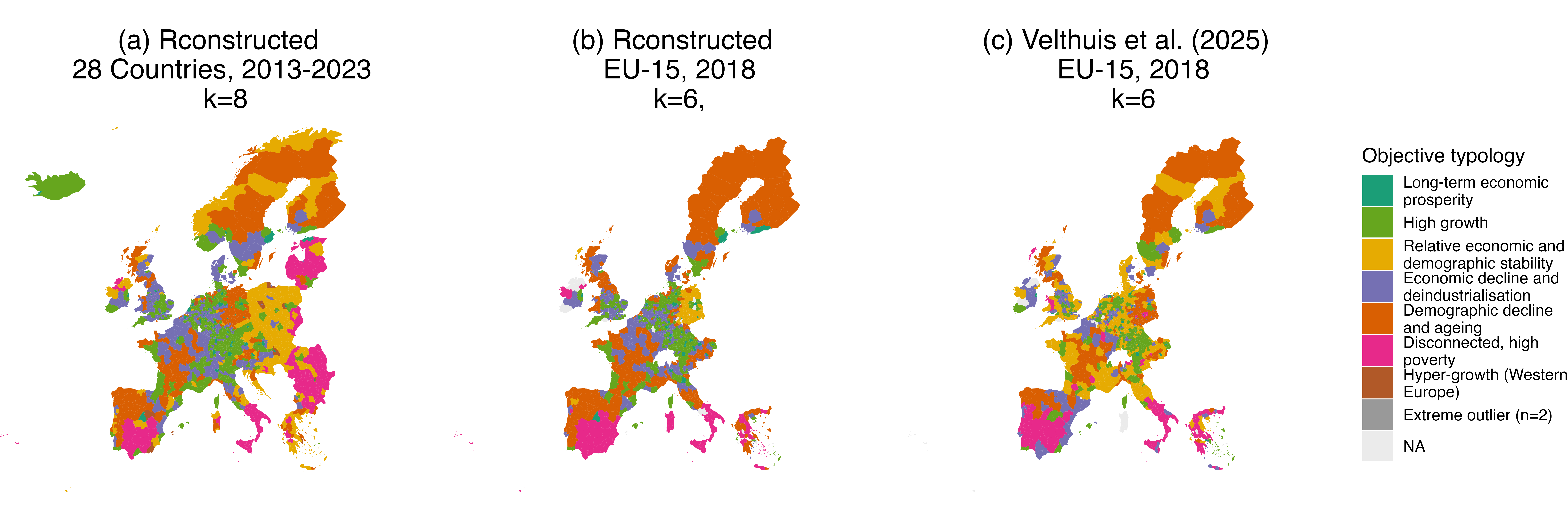}
\caption{The reconstructed objective left-behindness typology. Left to right: $k=8$ clusters (2013--2023), $k=6$ EU-15 (2018), and the published $k=6$ EU-15 (2018) typology of \cite{velthuis2025}.}
\label{fig:map-objective}
\end{figure}

Most profiles are reproduced. The principal divergence is per capita GDP growth for ``Relative economic and demographic stability'' ($+204.4$ versus $-5.0$ percentage points). This is not caused by cluster composition: in the published cluster, German regions show mildly positive growth ($+1.6$pp), while its seven East German regions show $-16.3$pp. The difference instead reflects later ARDECO revisions to growth estimates for these regions, particularly during the extreme post-reunification catch-up years. Figure~\ref{fig:map-objective} maps the resulting classifications.

\subsection{Composition: what each cluster looks like subjectively}
\label{sec:cluster-hfgi}

Table~\ref{tab:hfgi-profile} reports the mean and median of the three group
composites, Cohesion, Wellbeing and Illbeing, by cluster, for both typology
designs, computed directly from the raw, tweet-weighted, unadjusted $-1$ to
$+1$ HFGI scale (not residualized for urban-rural composition or tweet
volume). In both the main ($k=8$) and EU-15 benchmark ($k=6$) designs,
``Disconnected, high poverty'' has the \emph{highest} Cohesion, the
\emph{highest} Wellbeing and the \emph{lowest} Illbeing of any matched
cluster. ``Demographic decline and ageing'' has close to the opposite
Wellbeing profile in both designs; on Cohesion specifically it is,
essentially, tied with ``High growth'' for the lowest in the main design and
is clearly not the lowest in the EU-15 benchmark (``Relative economic and
demographic stability'' is). The poverty-disconnected periphery voices the
strongest cohesion \emph{and} the strongest personal well-being of any
objectively left-behind variety, and the demographically declining
periphery voices comparatively low well-being throughout, but is not
uniquely the lowest on the relational side.

\begin{table}[htbp]
\centering
\caption{HFGI composite mean/median by objective cluster (raw scale), both typology designs}
\label{tab:hfgi-profile}
\small
\resizebox{\textwidth}{!}{%
\begin{tabular}{lrrcccc}
\toprule
Cluster & $n$ & LB score & Composite & Cohesion & Wellbeing & Illbeing \\
\midrule
\multicolumn{7}{l}{\textit{Main typology, $k=8$, all 28 countries}} \\
Disconnected, high poverty & 120 & $+1.44$ & 0.609/0.517 & 0.650/0.592 & 0.522/0.442 & $-0.664$/$-0.741$ \\
Demographic decline and ageing & 219 & $+0.93$ & $-0.199$/$-0.220$ & $-0.149$/$-0.107$ & $-0.195$/$-0.277$ & 0.242/0.139 \\
Relative econ./demogr.\ stability & 160 & $+0.53$ & 0.130/$-0.044$ & 0.144/$-0.065$ & 0.123/0.042 & $-0.125$/0.201 \\
Economic decl.\ and deindustr. & 289 & $-0.13$ & $-0.075$/$-0.087$ & $-0.084$/$-0.072$ & $-0.052$/$-0.129$ & 0.091/0.106 \\
High growth & 305 & $-0.66$ & $-0.092$/$-0.140$ & $-0.146$/$-0.144$ & $-0.091$/$-0.167$ & 0.050/0.071 \\
Long-term economic prosperity & 109 & $-1.40$ & $-0.062$/$-0.185$ & $-0.055$/$-0.155$ & $-0.073$/$-0.261$ & 0.058/0.077 \\
Hyper-growth (Western Europe) & 13 & $-2.15$ & 0.017/0.216 & 0.028/0.246 & 0.059/0.261 & 0.033/0.004 \\
Extreme outlier & 2 & $-5.79$ & 0.858/0.858 & 1.007/1.007 & 1.210/1.210 & $-0.387$/$-0.387$ \\
\addlinespace
\multicolumn{7}{l}{\textit{EU-15 benchmark, $k=6$ (Step 1)}} \\
Relative econ./demogr.\ stability & 61 & $+1.67$ & $-0.214$/$-0.207$ & $-0.322$/$-0.334$ & $-0.323$/$-0.257$ & 0.019/0.034 \\
Disconnected, high poverty & 61 & $+1.09$ & 0.370/0.407 & 0.387/0.421 & 0.318/0.329 & $-0.409$/$-0.371$ \\
Demographic decline and ageing & 233 & $+0.80$ & $-0.219$/$-0.220$ & $-0.146$/$-0.081$ & $-0.184$/$-0.234$ & 0.314/0.164 \\
Economic decl.\ and deindustr. & 249 & $-0.30$ & $-0.124$/$-0.133$ & $-0.125$/$-0.059$ & $-0.097$/$-0.188$ & 0.151/0.126 \\
High growth & 227 & $-0.62$ & $-0.139$/$-0.179$ & $-0.233$/$-0.225$ & $-0.139$/$-0.238$ & 0.063/0.095 \\
Long-term economic prosperity & 88 & $-1.60$ & $-0.124$/$-0.213$ & $-0.129$/$-0.194$ & $-0.175$/$-0.300$ & 0.070/0.079 \\
\bottomrule
\end{tabular}%
}
\end{table}

One genuine nuance the median exposes that the mean alone would not: the
main typology's ``Relative economic and demographic stability'' cluster has
a Composite \emph{mean} of $+0.130$ but a \emph{median} of $-0.044$. This is a
right-skewed distribution consistent with this being the same East-German
catch-up cluster documented in Table~\ref{tab:table2}, where a handful of
extreme values pull the mean above zero while most member regions sit below
it. The EU-15 $k=6$ version of the same-named cluster shows a smaller
mean/median gap and is, on that typology, the single \emph{lowest}-cohesion
cluster of the six, a different position in the ranking than its $k=8$
mid-table placement. This is a reminder that ``Relative stability'' should
not be read as subjectively neutral on the strength of its name alone.

\subsection{Temporal evolution}
\label{sec:cluster-evolution}

\begin{figure}[htbp]
\centering
\includegraphics[width=0.8\textwidth]{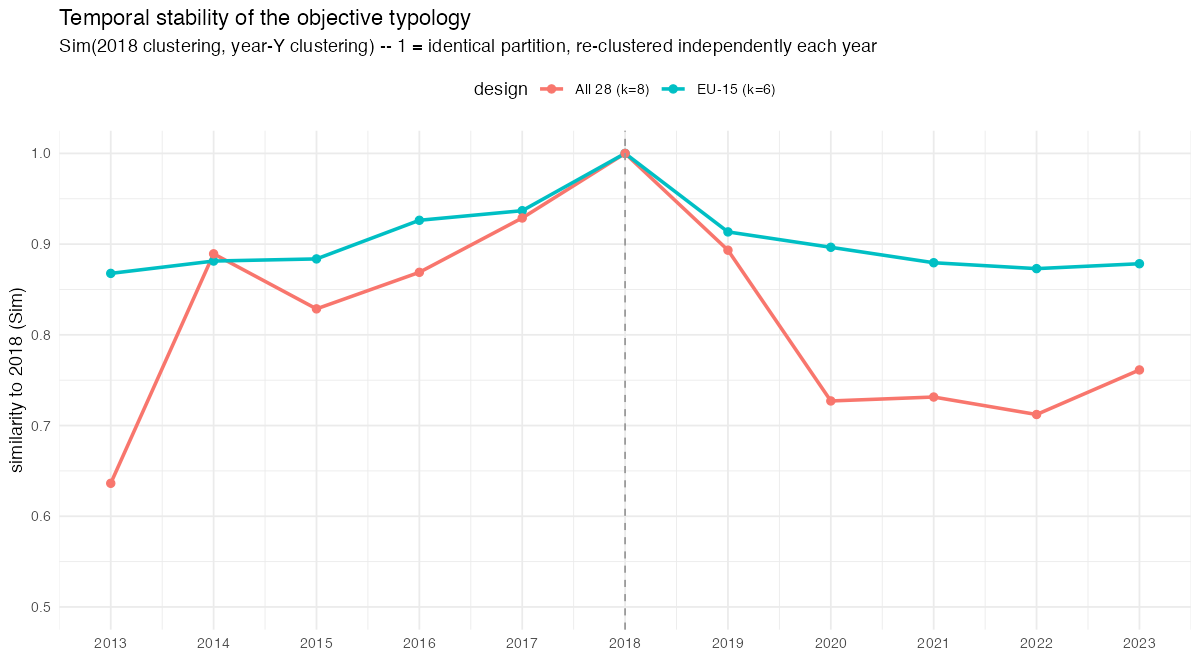}
\caption{Temporal stability of the objective typology: similarity to the 2018 reference partition, both designs, 2013--2023}
\label{fig:cluster-divergence}
\end{figure}

Cluster membership is moderately but not perfectly stable over 2013--2023.
Re-clustering each year independently (own standardization per year, same
$k$ and seed) and measuring each year's overlap with the fixed 2018 partition
via the Dice/S{\o}rensen group-similarity measure
\citep{dice1945,sorensen1948},
\begin{equation}
\mathrm{Sim}(g_1, g_2) = \frac{1}{|G_2|}\sum_{j \in G_2} \max_{i \in G_1}
\frac{2\,|\{n : g_1(n) = i\} \cap \{n : g_2(n) = j\}|}
     {|\{n : g_1(n) = i\}| + |\{n : g_2(n) = j\}|},
\end{equation}
similarity to the 2018 reference partition falls immediately after 2018 and
plateaus rather than continuing to decline (Figure~\ref{fig:cluster-divergence}):
for the main $k=8$ design, similarity is 0.64 in 2013, exactly 1.0 at the 2018
reference year by construction, and settles in the 0.71--0.76 range for
2020--2023; for the EU-15 $k=6$ design the pattern is similar, plateauing
around 0.87--0.93 after 2018. 

\begin{figure}[htbp]
\centering
\includegraphics[width=0.85\textwidth]{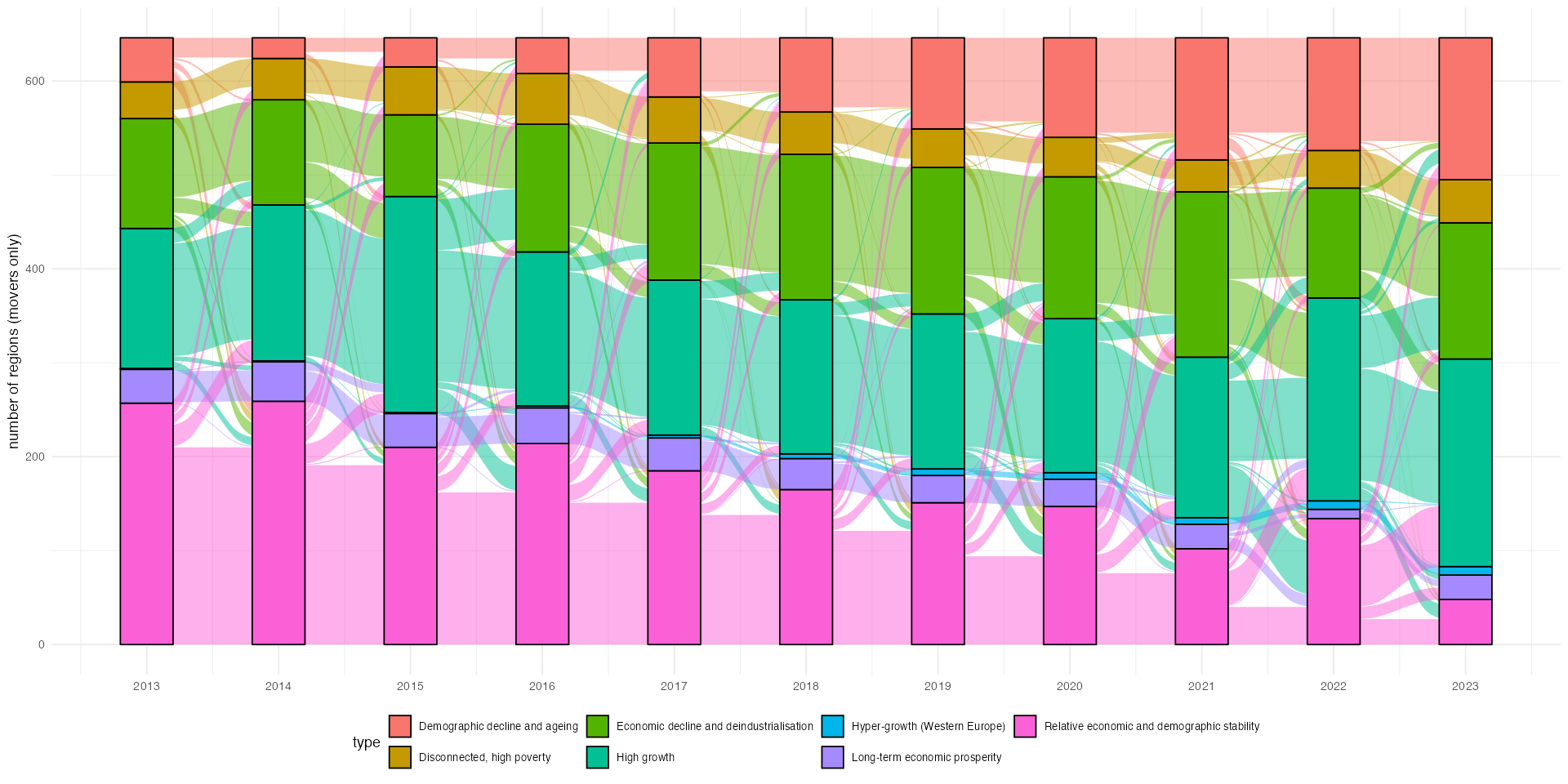}
\caption{Regions that changed left-behindness type at least once, 2013--2023, main $k=8$ design: only the 646 of 1,199 regions (53.9\%) that changed type at least once are shown (movers only; stable regions omitted for legibility); bar heights are mover counts per year}
\label{fig:alluvial-k8}
\end{figure}

Figure~\ref{fig:alluvial-k8} shows the regions that moved from one cluster
to another at least once over 2013--2023. 
The dominant pattern is not a
uniform drift in one direction: ``Relative economic and demographic
stability,'' the largest starting cluster among movers (257 of 646 in
2013), empties out almost completely by 2023 (48 remaining), with its
former members splitting in two opposite directions, 92 into
``Demographic decline and ageing'' and 72 into ``High growth.'' 
This bifurcation suggests that ``Relative stability'' is not a stable type
in its own right, but a transitional one: over time, its member regions
tend to move toward either demographic decline and ageing, or stronger
economic growth. ``Demographic decline and ageing'' is the largest net
gainer overall ($+104$). See Table~\ref{tab:mover-net-flow}.

\begin{table}[htbp]
\centering
\caption{Net cluster membership change among movers, 2013 vs.\ 2023, main $k=8$ design}
\label{tab:mover-net-flow}
\begin{tabular}{lrrr}
\toprule
Cluster & $n$ (2013) & $n$ (2023) & Net change \\
\midrule
Demographic decline and ageing & 47 & 151 & $+104$ \\
High growth & 149 & 221 & $+72$ \\
Economic decline and deindustrialisation & 117 & 145 & $+28$ \\
Hyper-growth (Western Europe) & 1 & 9 & $+8$ \\
Disconnected, high poverty & 39 & 46 & $+7$ \\
Long-term economic prosperity & 36 & 26 & $-10$ \\
Relative economic and demographic stability & 257 & 48 & $-209$ \\
\bottomrule
\multicolumn{4}{l}{\footnotesize $n$ counts are among the 646 of 1,199 regions (53.9\%) that changed cluster at least once, 2013--2023.} \\
\end{tabular}
\end{table}

\subsection{Do objective and subjective geographies diverge?}
\begin{figure}[p]
\centering
\includegraphics[width=\textwidth]{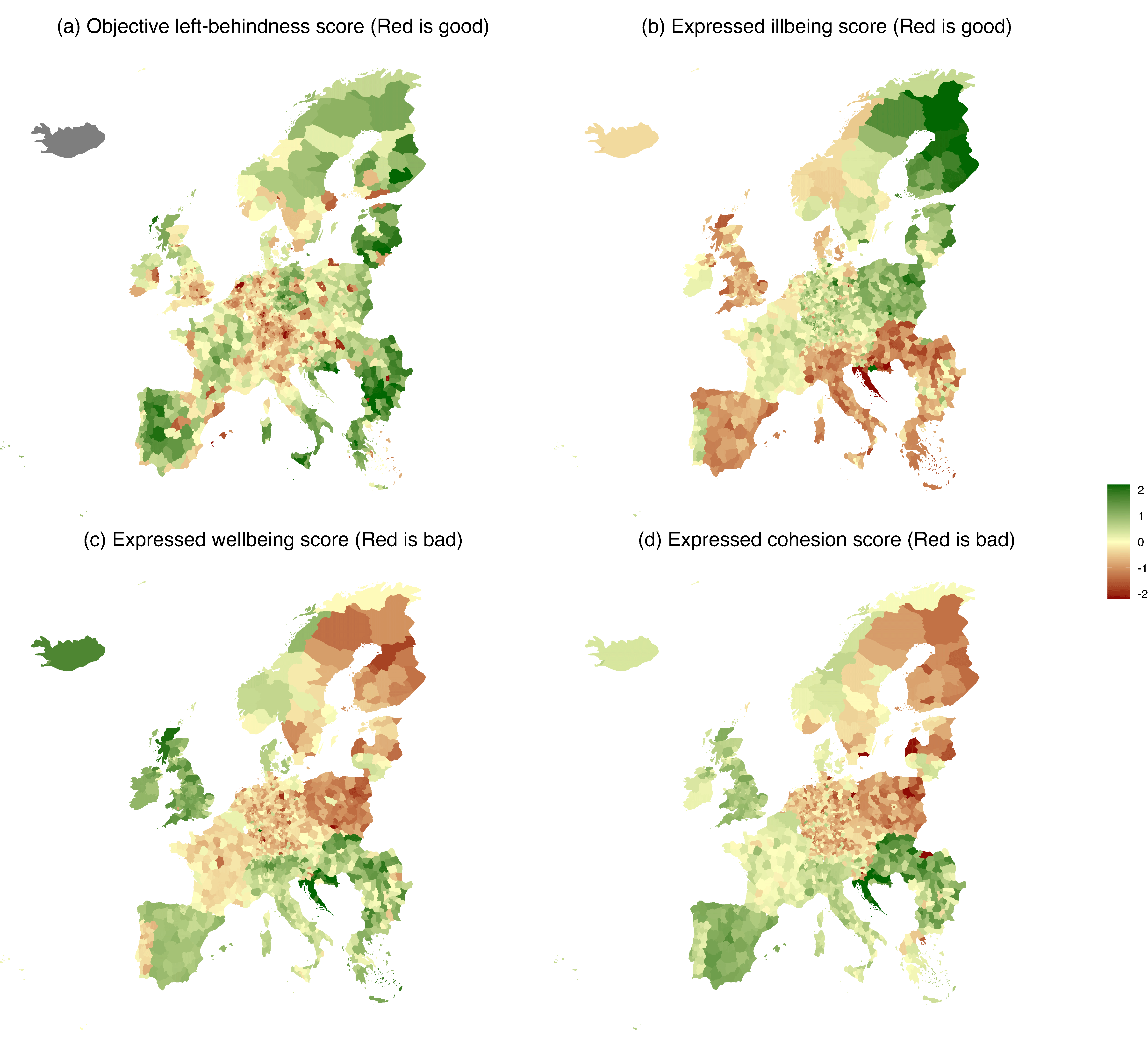}
\caption{Objective left-behindness score against expressed Cohesion, Well-being, and Ill-being, EU+UK+EFTA NUTS-3, 2018}
\label{fig:map-divergence}
\end{figure}

Figure~\ref{fig:map-divergence} shows an example of partial  spatial divergence of objective ($LB$ score) and subjective composite indicators. 
All four maps refer to the year 2018 and share one $z$-score colour scale (clipped at $\pm 2.2$) so
that colour intensity is directly comparable across panels. For the wellbeing (Panel c) and cohesion (Panel d) composite indicators, green marks a high degree of thriving and social bond while red marks less favorable conditions. The reverse is true for the objective left-behindness score (Panel a) and expressed illbeing (Panel b), i.e., green indicates poor outcomes. Visually, these spatial patterns suggest greater left-behindness in Northern, Eastern and Southern peripheral regions. However, the implications tend to diverge, regions left behind appear to be more distressed and less cohesive in Northern and Eastern peripherals, while the
opposite is true in the Southern part of Europe. 
To assess the extent and sources of this divergence more systematically, we next use a within--between modelling framework.

\subsection{Decomposing objective--subjective divergence through within--between modelling}
\label{sec:model}
 
The data covers 1,215 NUTS-3 regions and 12,864 region-year observations over 2013--2023 in an unbalanced panel. This panel structure allows us to ask two questions that a cross-sectional design cannot separate. 

\smallskip
\textbf{Question A}: \emph{do
regions that are, on average over the whole period, more objectively
left-behind than other regions also express, on average, different levels
of well-being, trust or belonging?} This is a comparison \emph{across}
regions, a geography-of-disadvantage question. 

\smallskip
\textbf{Question B}: \emph{when a
given region \emph{becomes} more left-behind than its own usual level, a
good year turning bad or vice versa, does that same region's own expressed
well-being move with it?} This is a comparison \emph{within} a region over
time, a change question rather than a geography question.

A pooled regression that ignores the panel structure conflates these two
questions into one coefficient, and the two not necessarily point the same way: a
region can be permanently more left-behind than its neighbours (a \emph{between}
fact) while also becoming somewhat less left-behind compared to its own past self
in a good year (a \emph{within} fact), and the effect of each on subjective
expression can differ in size or even in sign. A fixed-effects model would
isolate the within (over-time) effect cleanly, but at the cost of discarding
the between-region comparison entirely. The random-effects within-between (REWB)
model of \citet{bell2015} is designed for precisely this situation: it
decomposes a single time-varying regressor into its \emph{within-region} and
\emph{between-region} components within the same equation, retaining both,
while still allowing a random region intercept (rather than a fixed effect)
so that time-invariant regional characteristics are captured without
consuming one parameter per region.

Two specifications of this framework are estimated: \textbf{Model A}, using
the single aggregate left-behindness score, and \textbf{Model B}, using its
own ten underlying components instead. Both are instances of the same
general template below, and are compared directly in
Section~\ref{sec:within-between}.

For a vector of time-varying regressors $\mathbf{x}_{it}$ (for Model A,
the scalar left-behindness score $\mathrm{LB}_{it}$; for Model B, the
9-vector of the score's own raw, time-varying ingredients,
Table~\ref{tab:variable-sources}, observed at year $t$), define the
region's own mean over the panel, $\bar{\mathbf{x}}_i = T_i^{-1}\sum_t
\mathbf{x}_{it}$, and the deviation from that mean, $(\mathbf{x}_{it} -
\bar{\mathbf{x}}_i)$. Let $\mathbf{x}'_i$ denote any additional, purely
time-invariant regressors, with no within-region variation to decompose:
empty for Model A, and the score's tenth component, youth net migration,
for Model B. The model regresses the outcome on all these terms
simultaneously without country fixed effects:
\begin{align}
y_{it} = \beta_0
&+ \underbrace{\boldsymbol\beta_W^\top\,(\mathbf{x}_{it} - \bar{\mathbf{x}}_i)}_{\text{within (over time)}}
+ \underbrace{\boldsymbol\beta_B^\top\,\bar{\mathbf{x}}_i}_{\text{between (cross-national)}} \notag \\
&+ \boldsymbol\eta^\top \mathbf{x}'_i
+ \boldsymbol\gamma^\top \mathbf{v}_{it}
+ \boldsymbol\theta^\top \mathrm{URBN}_i
+ u_i + \varepsilon_{it},
\label{eq:rewb}
\end{align}
and with country fixed effects added to the contextual part:

\begin{align}
y_{it} = \beta_0
&+ \underbrace{\boldsymbol\beta_W^\top\,(\mathbf{x}_{it} - \bar{\mathbf{x}}_i)}_{\text{within (over time)}}
+ \underbrace{(\boldsymbol\beta_B^{\text{FE}})^\top\,\bar{\mathbf{x}}_i}_{\text{between (within-country)}}  + \boldsymbol\delta^\top \mathrm{COUNTRY}_{c(i)}
\notag \\
&+ \boldsymbol\eta^\top \mathbf{x}'_i
+ \boldsymbol\gamma^\top \mathbf{v}_{it}
+ \boldsymbol\theta^\top \mathrm{URBN}_i
+ u_i + \varepsilon_{it}.
\label{eq:rewb-fe}
\end{align}

In both equations, $u_i \sim N(0, \sigma_u^2)$ is a region-level random intercept
and $\varepsilon_{it} \sim N(0, \sigma_\varepsilon^2)$ the residual; $\mathbf
v_{it}$ is a vector of time-varying controls (here, $\log$ tweet volume).
The intraclass correlation $\rho = \sigma_u^2/(\sigma_u^2 +
\sigma_\varepsilon^2)$ is the share of residual variance in expressed
outcomes that is a stable, region-level characteristic rather than
year-to-year noise; a low $\rho$ means regions are not very persistently
different from one another on that outcome once the modelled regressors
are accounted for. Critically, $\boldsymbol\beta_W$ need not equal either
between coefficient: the model does not assume the within and between
effects are the same, and testing whether they differ is itself
informative about whether a variable's association with subjective
expression is a fact about \emph{places} (spatial, between-region) or a
fact about \emph{change} (temporal, within-region).

For Model A, where $\mathbf{x}_{it}$ is the scalar $\mathrm{LB}_{it}$,
$\boldsymbol\beta_W$ and $\boldsymbol\beta_B$ reduce to the scalars
$\beta_W$ and $\beta_B$, with a direct reading. $\beta_W$ is the within,
``over time'' coefficient, and the answer to Question B: \emph{if this same
region's own left-behindness rose by one standard deviation relative to
its own typical level, how much would its own expressed (HFGI) outcome
change?} Two regions can share the same average left-behindness over the
whole panel, yet, in a given year, sit on opposite sides of that average,
one currently \emph{above} it and the other \emph{below}; $\beta_W$ is
what each region's own deviation, $(x_{it}-\bar x_i)$, is multiplied by.
It is estimated purely from over-time variation within each region, so it
cannot be confounded by any time-invariant factors within a
region, such as language, national culture, or fixed geographical features, though it
can still be confounded by other things that change over time alongside
left-behindness.

$\beta_B$, the cross-national between coefficient in
equation~\eqref{eq:rewb}, answers Question A: \emph{comparing two
different regions, possibly in different countries, whose average
left-behindness differs by one standard deviation, how much does their
average expressed (HFGI) outcome differ?} Because most of the variance in left-behindness is between rather than within countries, this coefficient can absorb national-level differences that are unrelated to regional left-behindness itself, including institutional, economic, cultural, linguistic, and social-media expression differences. Adding country fixed
effects as in equation~\eqref{eq:rewb-fe} forces the same comparison to
happen \emph{within} each country instead, turning $\beta_B$ into
$\beta_B^{\text{FE}}$: two regions in the \emph{same} country that differ
in average left-behindness. This absorbs time-invariant country-level differences by construction, at the cost of discarding the larger cross-national contrast. A substantial change in the between coefficient from $\beta_B$ to $\beta_B^{\text{FE}}$ therefore indicates that the cross-national association is strongly conditioned by country-level differences; it does not identify which specific national mechanism is responsible.

Model A uses the aggregate score $\mathrm{LB}_{it}$ as $\mathbf{x}_{it}$
directly; Model B uses its ten components
instead.  Both models are estimated separately for each of the 14 raw
HFGI dimensions and the four composite indicators, 18 outcomes in total,
each fit both without and with country fixed effects. All variables are
standardized before estimation, so coefficients are in outcome
standard-deviation units per standard deviation of the regressor.

\section{Results}
\label{sec:within-between}

The estimation sample comprises 1,215 NUTS-3 regions and
12,864 region-year observations over 2013--2023. The panel is unbalanced:
1,091 regions are observed for all 11 years, 119 for seven years, and five
for six years.

Across the 36 within-between fits estimated for each model,
the region-level random-effect variance is estimated at or near zero in
9 to 15 specifications, depending on the model. The fixed-effect
coefficients are nevertheless obtained from models that converged normally
and are unaffected by this boundary estimate.

\subsection{Model A: the aggregate score}
\label{sec:results-1a}

Table~\ref{tab:model1a} reports, for every outcome, $\beta_B$
(cross-national), $\beta_B^{\text{FE}}$ (within-country), and $\beta_W$
(over time).

\begin{table}[htbp]
\centering
\caption{Model A: effect of the continuous left-behindness score on each subjective outcome (standard-deviation units)}
\label{tab:model1a}
\small
\begin{tabular}{lccc}
\toprule
Outcome & Cross-national & Within-country & Within \\
 & ($\beta_B$) & ($\beta_B^{\text{FE}}$) & ($\beta_W$) \\
\midrule
happiness & +0.034$^{*}$ & -0.033$^{**}$ & +0.027$^{***}$ \\
lifesat & +0.063$^{***}$ & -0.030$^{**}$ & +0.027$^{***}$ \\
anxiety & -0.010 & +0.007 & -0.002 \\
depression & +0.001 & +0.013 & -0.013 \\
belonging & +0.031$^{**}$ & -0.039$^{***}$ & +0.012 \\
discrim & -0.088$^{***}$ & -0.007 & +0.050$^{***}$ \\
govapprove & +0.082$^{***}$ & +0.003 & -0.070$^{***}$ \\
hope & +0.066$^{***}$ & -0.016 & -0.001 \\
loneliness & -0.008 & -0.002 & -0.042$^{***}$ \\
optimism & +0.035$^{*}$ & -0.035$^{**}$ & +0.029$^{***}$ \\
resilience & +0.005 & -0.051$^{***}$ & +0.058$^{***}$ \\
trust & +0.071$^{***}$ & -0.024$^{*}$ & -0.017$^{*}$ \\
trusted & +0.030$^{*}$ & +0.003 & -0.031$^{***}$ \\
fearfuture & -0.019 & +0.030$^{**}$ & +0.022$^{**}$ \\
composite & +0.047$^{***}$ & -0.023$^{*}$ & +0.001 \\
composite\_cohesion & +0.061$^{***}$ & -0.016 & -0.030$^{***}$ \\
composite\_wellbeing & +0.048$^{**}$ & -0.039$^{***}$ & +0.032$^{***}$ \\
composite\_illbeing & -0.029$^{*}$ & +0.010 & +0.003 \\
\bottomrule
\multicolumn{4}{l}{\footnotesize $^{*}p<.05$, $^{**}p<.01$, $^{***}p<.001$} \\
\end{tabular}
\end{table}

\paragraph{Cross-nationally, the cohesion pattern holds with zero sign
exceptions.} All nine well-being dimensions are positively signed (eight
significant; only resilience is null). All five ill-being dimensions are
negatively signed (only discrimination significant). More left-behind
regions express more well-being and cohesion, less ill-being, never the
reverse. All three group composites confirm this independently: Cohesion
($+0.061$, the strongest of the three), Wellbeing ($+0.048$), Illbeing
($-0.029$).

\paragraph{Country dummies reverse the well-being dimensions specifically,
and the reversal is concentrated in the personal-psychological group.}
Eight of nine well-being dimensions flip negative once country dummies are
added (six significantly: \emph{happiness, lifesat, belonging, optimism,
resilience, trust}); only \emph{govapprove} and \emph{trusted} stay flat.
Ill-being dimensions mostly stay null (\emph{fearfuture} is the one
exception, flipping positive). The within-country gradient is driven by
the well-being side reversing, not a symmetric reversal across every
dimension. Splitting Cohesion from Wellbeing shows exactly where:
Wellbeing reverses significantly ($-0.039$), driven by four of its five
dimensions; Cohesion's reversal is a wash ($-0.016$, n.s.), since only two
of its four dimensions flip and \emph{govapprove}/\emph{trusted} cancel
them out. Illbeing stays null ($+0.010$). The national-expression-norm
confound is a personal-well-being phenomenon, not a general one.

\paragraph{Within region, the pattern is genuinely mixed for individual
dimensions, but Cohesion and Wellbeing diverge cleanly once split.} Ten of
14 individual dimensions are significant over time, split across both
signs within well-being (\emph{happiness, lifesat, optimism, resilience}
rise; \emph{govapprove, trust, trusted} fall) and within ill-being
(\emph{discrim, fearfuture} rise; \emph{loneliness} falls). The pooled
Composite stays null ($+0.001$): too heterogeneous to survive pooling.
Splitting Cohesion from Wellbeing uncovers exactly the offsetting
relationship the two composites were built to expose
(Section~\ref{sec:data-subjective}): Cohesion falls ($-0.030$) while
Wellbeing rises ($+0.032$), both significant, cancelling into the null the
pooled Composite shows. Illbeing stays null ($+0.003$).

\subsection{Model B: individual objective measurements}
\label{sec:results-1b}

Table~\ref{tab:model1b-summary} compares, for every outcome, Model A's
aggregate score effect (cross-national) with Model B's single strongest
component at the same comparison. Every outcome has at least one
significant component, always larger, often two to three times larger,
than the aggregate score's own coefficient. The aggregate index dilutes a
real, concentrated effect by averaging it with nine components that do not
move the same way.

\begin{table}[htbp]
\centering
\caption{Model A's aggregate score vs.\ Model B's strongest individual component (cross-national, no country dummies)}
\label{tab:model1b-summary}
\small
\begin{tabular}{lccc}
\toprule
Outcome & LB score (A) & Strongest component (B) & Component effect \\
\midrule
happiness & +0.034$^{*}$ & Population growth differential & +0.079$^{***}$ \\
lifesat & +0.063$^{***}$ & Population growth differential & +0.112$^{***}$ \\
anxiety & -0.010 & At-risk-of-poverty rate & -0.087$^{***}$ \\
depression & +0.001 & At-risk-of-poverty rate & -0.080$^{***}$ \\
belonging & +0.031$^{**}$ & Net migration rate & -0.093$^{***}$ \\
discrim & -0.088$^{***}$ & Accessibility index & +0.102$^{***}$ \\
govapprove & +0.082$^{***}$ & Accessibility index & -0.099$^{***}$ \\
hope & +0.066$^{***}$ & Industrial employment share change & -0.124$^{***}$ \\
loneliness & -0.008 & At-risk-of-poverty rate & -0.105$^{***}$ \\
optimism & +0.035$^{*}$ & Population growth differential & +0.082$^{***}$ \\
resilience & +0.005 & Old-age dependency ratio & -0.153$^{***}$ \\
trust & +0.071$^{***}$ & Accessibility index & -0.086$^{***}$ \\
trusted & +0.030$^{*}$ & At-risk-of-poverty rate & +0.092$^{***}$ \\
fearfuture & -0.019 & At-risk-of-poverty rate & -0.089$^{***}$ \\
composite & +0.047$^{***}$ & Industrial employment share change & -0.078$^{***}$ \\
composite\_cohesion & +0.061$^{***}$ & Industrial employment share change & -0.075$^{***}$ \\
composite\_wellbeing & +0.048$^{**}$ & Accessibility index & -0.101$^{***}$ \\
composite\_illbeing & -0.029$^{*}$ & At-risk-of-poverty rate & -0.091$^{***}$ \\
\bottomrule
\multicolumn{4}{l}{\footnotesize $^{*}p<.05$, $^{**}p<.01$, $^{***}p<.001$} \\
\end{tabular}
\end{table}

Two patterns stand out. \textbf{At-risk-of-poverty rate} is the strongest
component for six outcomes (\emph{anxiety, depression, loneliness,
trusted, fearfuture, composite\_illbeing}), always worsening ill-being,
and, via \emph{trusted}, raising a well-being dimension too.
\textbf{Accessibility} is the strongest component for
\emph{discrimination}, \emph{government approval}, \emph{trust}, and
\emph{composite\_wellbeing}: \emph{better} accessibility means
\emph{lower} government approval, trust, and personal well-being, and
\emph{more} discrimination, the opposite of a naive ``isolation is bad''
story, and something the aggregate score's own positive coefficients
(Table~\ref{tab:model1a}) do not reveal. \textbf{Industrial employment
share change} is the strongest component for \emph{hope}, the pooled
\emph{composite}, and \emph{composite\_cohesion}: once Cohesion is
isolated from Wellbeing, de-industrialization, not accessibility, carries
its score effect.

Full coefficients for all ten components, all 18 outcomes, are in
Tables~\ref{tab:m1b-happiness}--\ref{tab:m1b-compositeillbeing} in the Supplementary Material. Two
patterns recur. \textbf{Old-age dependency} has by far the largest within
coefficient for almost every well-being dimension ($-0.500$ for the pooled
Composite, $-0.617$ for government approval, $-0.538$ for trust), an order
of magnitude larger than any other within effect. A region's own rising
old-age dependency is the single most consequential change for its own
well-being. \textbf{Industrial employment share change} diverges:
cross-nationally, more de-industrialization means \emph{higher} well-being
and cohesion; within region over time, a region's \emph{own}
de-industrialization means \emph{lower} well-being. Same component,
opposite signs depending on whether the comparison is across places or
across time, exactly the distinction Section~\ref{sec:model}
motivates the within-between decomposition to make visible.

Yet the within effects reported so far are average temporal associations. If the consequences of territorial decline depend on local institutional capacity, social structure or development trajectories, these averages may conceal substantial regional variation in both the strength and the direction of the relationship (See Section~\ref{sec:rand_slp}).

\subsection{Nonlinearities in the relationship between territorial disadvantage and subjective outcomes}
\label{sec:model-h3}
The relationship between territorial disadvantage and subjective outcomes is not necessarily linear. In particular, belonging and cohesion may persist or even increase under moderate levels of disadvantage before weakening at more severe levels. We examine this possibility for all 18 outcomes using a region-level cross-sectional quadratic regression, with one observation per NUTS-3 region obtained by averaging over the panel:
\begin{equation}
y_i = a + b_1\,\mathrm{LB}_i + b_2\,\mathrm{LB}_i^2
+ \boldsymbol\theta^\top\mathrm{URBN}_i + \boldsymbol\delta^\top\mathrm{COUNTRY}_{c(i)} + e_i.
\label{eq:quadratic}
\end{equation}
A significant $b_2$ is not, by itself, evidence of a genuine hump or trough:
the vertex of the fitted parabola, $-b_1/2b_2$, must additionally fall
inside the \emph{central} 80\% of the observed left-behindness distribution
(10th--90th percentile), not merely between its sample minimum and maximum,
otherwise curvature driven by a handful of extreme regions would be misread
as a real turnaround. With 18 simultaneous tests, a
Benjamini--Hochberg false-discovery-rate-adjusted $p$-value is reported
alongside the raw one.
Table~\ref{tab:h3} reports the quadratic term, its raw and FDR-adjusted
$p$-value, and the shape classification according to equation~\ref{eq:quadratic} for
each outcome, ordered by  significance.

\begin{table}[htbp]
\centering
\caption{Nonlinearity tests for the relationship between left-behindness and subjective outcomes, all 18 outcomes}
\label{tab:h3}
\small
\begin{tabular}{lrrr>{\raggedright\arraybackslash}p{6cm}}
\toprule
Outcome & $b_2$ & raw $p$ & FDR $p$ & Shape \\
\midrule
{belonging} & +0.031 & 0.0007 & 0.0121 & significant curvature, vertex in the tail decile (n=26 regions beyond it) -- not a robust hump/trough \\
{trusted} & +0.029 & 0.0095 & 0.0583 & TRUE U-shape (falls then rises) -- does not survive FDR correction \\
{composite\_cohesion} & +0.025 & 0.0097 & 0.0583 & TRUE U-shape (falls then rises) -- does not survive FDR correction \\
{discrim} & -0.018 & 0.0321 & 0.1443 & significant curvature, vertex in the tail decile (n=87 regions beyond it) -- not a robust hump/trough \\
lifesat & +0.015 & 0.1218 & 0.4166 & linear \\
loneliness & -0.013 & 0.1403 & 0.4166 & linear \\
trust & +0.017 & 0.1620 & 0.4166 & linear \\
composite & +0.012 & 0.1955 & 0.4398 & linear \\
composite\_illbeing & -0.011 & 0.2514 & 0.5023 & linear \\
depression & -0.011 & 0.2791 & 0.5023 & linear \\
happiness & +0.012 & 0.3138 & 0.5078 & linear \\
resilience & -0.012 & 0.3631 & 0.5078 & linear \\
hope & -0.008 & 0.3667 & 0.5078 & linear \\
optimism & +0.009 & 0.4783 & 0.6149 & linear \\
govapprove & +0.006 & 0.5671 & 0.6805 & linear \\
composite\_wellbeing & +0.003 & 0.7664 & 0.8142 & linear \\
anxiety & -0.003 & 0.7689 & 0.8142 & linear \\
fearfuture & -0.001 & 0.9219 & 0.9219 & linear \\
\bottomrule
\end{tabular}
\end{table}

No dimension or composite shows a robust hump or trough once the
vertex-location and false-discovery-rate checks  are applied. Belonging is the one outcome whose
quadratic term survives FDR correction ($q=0.012$), but its vertex is
supported by only 26 of 1,369 regions. This is curvature driven by a tail of
extreme regions, not a genuine turnaround, and is formally not classified as
a true inverted-U. \emph{Trusted}, \emph{discrimination} and, once split
from the pooled Cohesion/Wellbeing group, \emph{composite\_cohesion} itself
clear the raw $p<.05$ threshold but none survives FDR correction across the
18 simultaneous tests. Overall, we find no robust evidence of an inverted-U relationship between territorial disadvantage and any of the 18 subjective outcomes under this specification. The positive linear cross-national association between left-behindness and belonging is reported separately in Table~\ref{tab:model1a}; it should not be interpreted as evidence of a nonlinear resilience threshold.

\subsection{Heterogeneity in the within-region effect of left-behindness} 
\label{sec:rand_slp}
Regions may differ substantially in their sensitivity to changes in left-behindness. Similar levels of deterioration in economic and demographic conditions may have markedly different consequences depending on local institutional capacity, social capital, population composition, cultural and/or broader regional development trajectories. To investigate this possibility, we extended the baseline specification (Eq.~\eqref{eq:rewb}) by allowing the within-region coefficient on left-behindness to vary across NUTS-3 regions through a random-slope specification (Eq.~\eqref{eq:rewb-fe-rs}).

\begin{align}
y_{it} = \beta_0 
& + \underbrace{\boldsymbol(\beta_{W,LB}+b_i)^\top (LB_{it}-\bar{LB}_i)}_{\text{within (over time)}} + 
\underbrace{(\boldsymbol\beta_{B,LB})^\top \bar{LB}_i}_{\text{between (within-country)}} + 
\notag\\ 
& + \boldsymbol\eta^\top \mathbf{x}'_i + \boldsymbol\gamma^\top \mathbf{v}_{it} + \boldsymbol\theta^\top \mathrm{URBN}_i + u_i  + \varepsilon_{it}, 
\label{eq:rewb-fe-rs} 
\end{align} 
$$ \begin{pmatrix} u_i\\ b_i \end{pmatrix} \sim N \!\left( \begin{pmatrix} 0\\ 0 \end{pmatrix}, \begin{pmatrix} \sigma_u^2 & \sigma_{ub}\\ \sigma_{ub} & \sigma_b^2 \end{pmatrix} \right). $$ 

Table \ref{tab:lb_random_slopes} provides evidence of substantial heterogeneity in the relationship between left-behindness and subjective expressions. For nearly all outcomes, likelihood-ratio tests strongly reject a model containing only random intercepts in favour of one that additionally allows cross-region heterogeneity in the within left-behindness effect. The estimated standard deviation of the random slopes ranges from negligible values for resilience and loneliness to considerably larger values for composite outcomes, indicating that the consequences of increasing left-behindness are far from uniform across Europe.

The strongest evidence of spatial heterogeneity emerges for the composite indicators. For overall flourishing, the random-slope standard deviation reaches 0.701 and the likelihood-ratio statistic exceeds 240, suggesting substantial variation in how regional changes in left-behindness translate into broader well-being outcomes. Heterogeneity is even greater for the cohesion composite (SD = 0.893), while notable variation is also observed for the well-being (SD = 0.590) and ill-being composites (SD = 0.711). These findings imply that the social and psychological consequences of left-behindness are strongly conditioned by local contexts rather than being fixed across space.

\begin{table}[htbp] 
\centering \caption{Heterogeneity in the within-region effect of left-behindness score (random slope)} \label{tab:lb_random_slopes} \begin{tabular}{lccc} 
\toprule 
Outcome & Avg. within effect of LB & SD of LB slope & LR statistic \\ 
\midrule
loneliness & -0.001*** & 0.000 & 0.048\\
resilience & 0.002*** & 0.002 & 4.778\\
govapprove & -0.002*** & 0.008 & 291.650***\\
trusted & -0.002*** & 0.013 & 166.800***\\
discrim & 0.004*** & 0.024 & 373.176***\\
\addlinespace
hope & 0.003* & 0.026 & 226.407***\\
trust & -0.002 & 0.039 & 219.529***\\
fearfuture & 0.005* & 0.043 & 159.632***\\
lifesat & 0.012*** & 0.044 & 167.065***\\
anxiety & -0.003 & 0.068 & 222.278***\\
\addlinespace
depression & -0.009* & 0.070 & 134.129***\\
optimism & 0.020*** & 0.087 & 153.020***\\
belonging & 0.011* & 0.091 & 156.783***\\
happiness & 0.026*** & 0.107 & 120.372***\\
composite wellbeing & 0.184*** & 0.590 & 176.924***\\
\addlinespace
composite & 0.056 & 0.701 & 241.494***\\
composite illbeing & -0.019 & 0.711 & 191.133***\\
composite cohesion & -0.098* & 0.893 & 295.087***\\
\bottomrule 
\end{tabular} 

\vspace{0.5em} 
\footnotesize \textit{Notes:} The table reports the estimated mean random slope for the within-region left behind effect, the estimated standard deviation of the regional random-slope distribution, and the likelihood-ratio (LR) test comparing a random-intercept model (Eq~\eqref{eq:rewb}) with a random-intercept-and-slope model (Eq~\eqref{eq:rewb-fe-rs}). Significance markers are inherited from the original estimates: $^{*}p<0.05$, $^{**}p<0.01$, $^{***}p<0.001$. 
\end{table}

\begin{figure}[p]
\centering
\includegraphics[width=\textwidth]{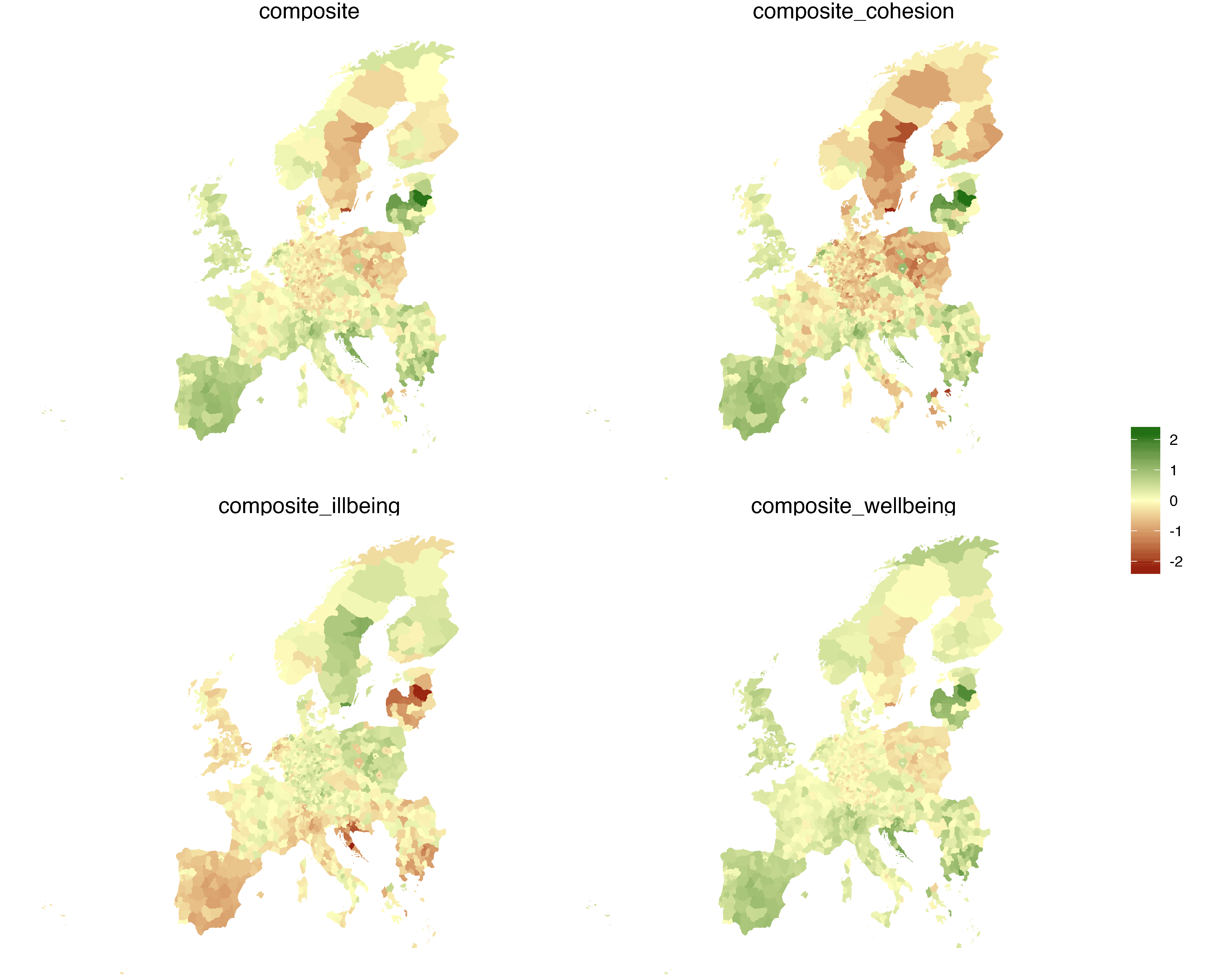}
\caption{Spatial heterogeneity of within-region effects of left-behindness score composite flourishing, EU+UK+EFTA NUTS-3. \textit{Notes:} Color corresponds to the estimates of $\beta_{W,LB}+b_i$ in Eq. \eqref{eq:rewb-fe-rs}. For visualization, color scale is clipped at the 80\% of maximum absolute value.} 
\label{fig:map-randslp-composite}
\end{figure}

The maps in Figure \ref{fig:map-randslp-composite} illustrate this spatial variation. Positive/negative values indicate regions where  increasing left-behindness is associated with growing/declining outcomes. Several broad geographical patterns are apparent.

For the overall flourishing composite, positive slopes are concentrated in parts of Southern Europe, including much of Spain, Italy, and Greece, as well as portions of the Balkans and the Baltic states. By contrast, negative slopes are more prevalent in parts of Northern and Central Europe, particularly Sweden and sections of Poland. Thus, changes in left-behindness appear to have substantially different implications for flourishing across European regional contexts.

The cohesion composite exhibits the clearest spatial concentration. Large negative slopes are observed across substantial parts of Sweden and parts of Eastern Europe, indicating that increasing left-behindness is associated with declining expressions of trust, belonging, and institutional confidence in these regions. Conversely, more positive slopes are found in the Baltic states and parts of Southeastern Europe (most notably Spain). This pattern suggests that social cohesion may be particularly sensitive to territorial disadvantage in some national and regional contexts while remaining comparatively resilient in others.

The well-being composite displays a different geography. Positive slopes are widespread across Southern Europe, the Balkans, and the Baltic region, whereas weaker or negative relationships appear across parts of Scandinavia and Central Europe. The spatial distribution indicates that the psychological dimensions of flourishing respond differently to left-behindness than the social-cohesive dimensions, reinforcing the rationale for analysing these components separately.

Finally, the ill-being composite reveals a reversed pattern. While some regions in Scandinavia and Eastern Europe display positive slopes, suggesting increases in anxiety, depression, loneliness, and related expressions alongside rising left-behindness, other regions show the opposite relationship. Similar mapping for each flourishing dimension is presented in Figure~\ref{fig:map-randslp-dim} in the Supplementary Material.

Taken together, these findings demonstrate that the consequences of left-behindness are not spatially uniform. Although the average within-region effects reported earlier in Table~\ref{tab:model1a} provide evidence of significant associations between left-behind conditions and expressed flourishing, the random-slope models reveal that the magnitude and even direction of these relationships vary markedly across European regions. This suggests that territorial disadvantage interacts with region-specific social, institutional, and developmental contexts. Consequently, policies aimed at mitigating left-behindness cannot assume identical effects across places; rather, the well-being and cohesion consequences of territorial decline appear to depend critically on local conditions and capacities.

\section{Conclusions}
\label{sec:conclusions}

This paper examines whether the material geography of territorial left-behindness corresponds to expressed well-being, ill-being and social cohesion across European regions. Combining ten years of economic, demographic and spatial indicators with geographically disaggregated social-media measures allows us to compare regions across Europe, within countries and within the same region over time.

The central finding is that these geographies correspond only partially. Regions with similar material disadvantages can display substantially different subjective profiles, while some highly disadvantaged profiles show relatively favourable well-being or cohesion. Left-behindness therefore has no single subjective correlate, extending evidence that its dimensions relate differently to life satisfaction and supporting its treatment as a multidimensional condition \citep{PanoriEtAl2025,FiorentinoEtAl2024}. The divergence is especially visible for cohesion. This does not imply that disadvantage strengthens social ties: economic and demographic decline can erode communities and social infrastructure \citep{FiorentinoEtAl2024}. Rather, material conditions, belonging, place attachment and local social capital are distinct dimensions of regional development \citep{MacKinnonEtAl2022}.

Geographical scale substantially changes the results. Associations that appear favourable in cross-country comparisons weaken or reverse within countries; for personal well-being, the association with left-behindness becomes negative after controlling for country differences. National economic, institutional, cultural and linguistic contexts, as well as differences in social-media expression, may therefore shape the European cross-section. Although our design cannot separate these mechanisms, it shows that subjective correlates cannot be interpreted independently of national context.

The temporal results further distinguish being disadvantaged from becoming more disadvantaged, consistent with accounts of left-behindness as cumulative relative decline \citep{FiorentinoEtAl2024}. Rising old-age dependency is associated with deterioration in several outcomes within regions, poverty more closely with psychological distress, and accessibility and deindustrialisation show different between- and within-region relationships. Within-region effects also vary substantially across regions and countries in both magnitude and direction. Moreover, many regions change typology over the decade, showing that cross-sectional categories can combine different trajectories.

Aggregate indices conceal much of this heterogeneity. The overall left-behindness score often has weak associations where individual components show stronger and more specific relationships; similarly, well-being, ill-being and cohesion can move in different directions and offset one another in a subjective composite. We also find little robust evidence that moderate disadvantage generally strengthens cohesion or belonging before deterioration at higher levels. High cohesion in some disadvantaged regions should therefore not be interpreted as a common resilience threshold.

The paper extends recent work linking objective left-behindness to subjective well-being \citep{PanoriEtAl2025} in three ways: through a ten-year panel that separates persistent regional differences from within-region change; broader, fine-grained European coverage; and subjective indicators derived from a large corpus of georeferenced public discourse rather than predefined survey questions. The resulting dataset combines harmonised objective and subjective indicators at NUTS-3 level, enabling left-behindness to be studied as a changing territorial process. HFGI measures expressed states among observed social-media users, not representative population attitudes, and should complement rather than replace survey and administrative evidence.

Overall, the subjective correlates of left-behindness depend on its form, national context and regional trajectory. Policies should therefore not treat regions with similar aggregate scores as facing equivalent problems. Combining objective indicators with well-being and cohesion measures can distinguish persistent disadvantage from ongoing deterioration and identify places where material and subjective conditions move in different directions.

\section*{Disclosure Statement}
No potential conflict of interest was reported by the
authors.

\section*{Acknowledgments}
Computational resources were provided by the National Academic Infrastructure for Supercomputing in Sweden (NAISS), funded by the Swedish Research Council.

\bibliographystyle{apalike}
\bibliography{biblio}

\clearpage
\appendix

\section*{Supplementary material}
\addcontentsline{toc}{section}{Supplementary material}

\setcounter{section}{0}
\renewcommand{\thesection}{S\arabic{section}}

\setcounter{table}{0}
\setcounter{figure}{0}
\renewcommand{\thetable}{S\arabic{table}}
\renewcommand{\thefigure}{S\arabic{figure}}

\section{Construction and harmonisation of territorial indicators}
\label{sec:supp-indicator-construction}
\paragraph{Ratios, differentials and levels.} To remain consistent with the reference framework, each variable is expressed in
the same form as the corresponding variable in \cite{velthuis2025}: GDP per capita relative to the national mean is a \emph{ratio}
(per capita $GDP_i$/per capita $GDP_{\text{nation}}$), not a raw level, so it measures
relative rather than absolute prosperity. GDP per capita, employment and population
growth are \emph{differentials}: a region's own growth rate \emph{minus} its
country's growth rate over the same window. This isolates whether a region
grew faster or slower than its own national trend, not merely whether it
grew. Industrial-employment-share change is a raw \emph{point change} with no
national term (a region's own de-industrialization, not a relative one). Net
migration, old-age dependency, poverty and accessibility are used as raw
\emph{levels}. The growth window runs from 1991 \emph{``given the importance of
long-term economic change in shaping perceptions of left-behindness,''} as \cite{velthuis2025} correctly signal. Industrial
employment share is benchmarked from 1995, the earliest year ARDECO's sectoral
series is available. Six countries: Estonia, Latvia, Lithuania, Croatia,
Slovakia, and Czechia lack pre-independence national
accounts. For those countries the growth baseline is its own earliest available
year rather than a single global 1991 baseline, which would otherwise drop
them entirely.

\paragraph{Youth migration.} \cite{velthuis2025} construct youth migration
directly from single-year-of-age population counts. Eurostat does not publish
single-year-of-age population at NUTS-3 level but only the five-year age bands are
available. We instead estimate it with the standard cohort-component method: the 15--19 and
20--24 age bands (\texttt{demo\_r\_pjangrp3}) are aged forward by one
band-width (five years, 2014$\rightarrow$2019, matching the reference paper
demographic window), and the residual between the aged cohort and the
observed later population is attributed to net migration (assuming mortality at these
ages is negligible, the standard simplifying assumption in youth-migration
studies), annualized to a per-1,000 rate on the same scale as the general net
migration rate. This is a single structural estimate for the whole panel, not
recomputed per year, matching the treatment of the underlying age structure
itself. It achieves complete coverage of the clustering sample and was
validated against \cite{velthuis2025} own published per-cluster youth-migration
figures before use: the same sign and the same relative ranking hold across
all six of their matched cluster types (Table~\ref{tab:table2}).

\paragraph{ESPON recovery.} Raw ESPON releases are internally inconsistent in
NUTS vintage: poverty and income are published on NUTS~2016 boundaries,
accessibility on NUTS~2010. Neither vintage joins directly to the NUTS~2021
panel used throughout this analysis, since NUTS boundaries were substantially
revised between vintages. Every ESPON code is first remapped to NUTS~2021 via
a centroid-in-polygon crosswalk between the relevant GISCO NUTS-boundary
vintages \citep{gisco}, recovering accessibility to 77--100\% of NUTS-3 regions per country
and poverty coverage across all 28 countries. Two genuine data gaps remain
after remapping: Belgium and the United Kingdom lack any ESPON poverty
observations, a coverage gap rather than a vintage mismatch. These are
filled with a spatial generalized additive model (GAM) \citep{wood2003thinplate,wood2017gam}. Let $\text{gdp}_{it}$ be the GDP per capita for region $i$ in year $t$, and $\text{lon}_i$, $\text{lat}_i$ be latitude and longitude of the centroid of region $i$. Then we fit a GAM model described by the following equation:
$$\text{poverty}_{it} = f(\text{lon}_i, \text{lat}_i) + \eta \log(\text{gdp}_{it})
+ \tau_t + \xi_{it},$$ 
with $f$ a thin-plate spline. The fitted model has an explained deviance of 0.71,
against 0.24 for a GDP-only model, confirming poverty is strongly spatially
structured net of income.  Observed values are kept wherever available and
predictions are used only where ESPON has no observation.
The model is validated out of sample: withholding each country in turn. For example, the
predicted national poverty rate for the UK is 14.9 against an observed
15.3, and for France 13.5 against an observed 13.8. We consider this as a reasonable approximation though not a perfect measurement.

\paragraph{Urban-rural classification.} In the statistical modeling of Section~\ref{sec:model} we also make use of the  Eurostat's
official urban-rural typology as a time-invariant control,
$\mathrm{URBN}_i \in \{$urban, intermediate, rural$\}$,
attached to each NUTS-3 region via GISCO's NUTS geometry \citep{gisco,giscoR}
attributes. We will use this to control for the possibility that urban and rural regions differ
systematically in both objective conditions and Twitter/X discourse
patterns (e.g.\ tweet volume, topic composition) in statistical modeling, independently of
left-behindness itself.

\section{Why the reconstruction diverges from the reference}
\label{sec:cluster-divergence-sources}

Although we restrict the comparison to the 15 countries studied by
\cite{velthuis2025} and reproduce their variable definitions as closely as
possible, exact region-level agreement across the six clusters is 47.3\%.
Several factors can account for this divergence.

\paragraph{Data vintage.}
ARDECO historical series are revised as national accounts are updated, and
our reconstruction uses a later vintage than that available to
\cite{velthuis2025}. The consequences can be substantial. For example,
Table~\ref{tab:table2} reports GDP growth of $+204.4$ percentage points for
the matched ``Relative economic and demographic stability'' cluster,
compared with $-5.0$ in \cite{velthuis2025}, despite the common 1991
baseline. Differences of this magnitude can materially affect cluster
assignment.

\paragraph{Differences in variable construction.}
Some variables cannot be reproduced identically. Poverty and accessibility
are based on harmonized NUTS-2021 ESPON series, with a spatial GAM used to
fill residual UK and Belgian gaps (Section~\ref{sec:data-objective}).
Likewise, youth migration is reconstructed from Eurostat five-year age
bands because single-year-of-age population data are unavailable at
NUTS-3, whereas \cite{velthuis2025} used single-year ages. These differences
affect a subset of regional values.

\paragraph{Standardization and clustering.}
Differences in the underlying data also alter the reference distribution
used for standardization and Mahalanobis outlier detection. They can
therefore propagate beyond the regions directly affected. Moreover,
$k$-means assignments can be sensitive to relatively small changes in the
input data, particularly for observations near cluster boundaries.

\paragraph{Nature of the disagreement.}
Importantly, disagreement is substantially smaller at the broader level:
binary left-behind/not-left-behind agreement is 76.1\%
(Cohen's $\kappa=0.524$), compared with 47.3\% exact agreement across the
six clusters. Thus, much of the divergence concerns \emph{which type} of
left-behind or non-left-behind region is identified, rather than whether a
region is classified as left behind at all.

\section{The GADM to NUTS crosswalk}\label{sec:crosswalk}

The subjective HFGI data are defined on GADM administrative boundaries
(version 2.8) \citep{gadm}, whereas the objective indicators are defined on
NUTS regions. We therefore constructed a crosswalk between the two
geographies. Our primary crosswalk was obtained by intersecting GADM~2.8
and NUTS~2021 polygons in an equal-area projection and assigning each GADM
unit to the NUTS region with the largest spatial overlap. As an independent
validation, we compared these assignments with the GADM$\leftrightarrow$NUTS
correspondence provided by the COVID-19 Data Hub
\citep{guidotti2020covid19}. The two methods agree on 96.4\% of
ADM2$\to$NUTS3 assignments for which both provide a match. Since the Data
Hub correspondence has more limited coverage at NUTS-3 level, we use the
spatial crosswalk throughout the analysis.

We use ADM2 to NUTS3 rather than ADM1 because the correspondence between
GADM administrative levels and NUTS levels varies across countries. For
example, ADM1 corresponds to NUTS2 in Italy, NUTS3 in Czechia, and entire
countries in the UK. When multiple ADM2 units map to the same NUTS-3
region, subjective indicators are aggregated using tweet-count-weighted
means. We also include the logarithm of region-year tweet volume as a
covariate in the models to account for differences in the amount of
information underlying each regional estimate.

\section{Potential biases and limitations of social media based indicators}\label{sec:issues}

Social media data are not a representative sample of the population and are subject to several forms of self-selection \citep{olteanu2019}. Twitter users tend to be younger, more educated, higher-income, and more urban than the general population \citep{wojcikhughes2019, mislove2011, hargittai2020}, while users who produce geolocated tweets are further selected geographically, particularly toward urban and coastal areas \citep{malik2015}. Our HFGI indicators should therefore not be interpreted as population estimates of belonging, trust, or well-being. They measure the prevalence of these expressions in the geolocated Twitter population observed in each territory.

Aggregation at the regional level mitigates some, but not all, consequences of this selection. Persistent differences in the demographic composition or propensity to use Twitter may affect the levels of the subjective indicators without necessarily explaining their spatial and temporal association with objective territorial disadvantage. More problematic would be differential selection: for example, if the composition of geolocated Twitter users, or their propensity to express belonging, trust, or well-being, changed systematically with local conditions. Such changes could generate apparent within-region variation unrelated to changes in the underlying population.

A related source of selection concerns expression itself. Individuals who publicly express positive or negative evaluations may differ systematically from those who remain silent, and these differences may be related to socioeconomic circumstances or other unobserved characteristics \citep{olteanu2019}. The use of the full geolocated tweet stream avoids selection into a dedicated survey or topic-specific channel, but it cannot eliminate selection into public expression. Consequently, we interpret the social-media indicators as measures of \emph{expressed subjective experience}, rather than representative measures of population attitudes. The objective--subjective relationships documented in this work should be read with this distinction in mind.

\section{Extracting Human-Flourishing Dimensions from Twitter/X}
\label{sec:howto}
HFGI was built using the large-language-model approach developed in
\cite{iacus2026} to identify expressions related to human flourishing
in geolocated Twitter/X posts. Each tweet is classified independently
across 46 dimensions, including the dimensions used in this study, such
as life satisfaction, belonging to society, trust, optimism, and other
aspects of subjective well-being.
For each dimension detected in a tweet, the classifier assigns one of
three intensity labels:
 \emph{low}: negative or opposite expression of the construct;
    \emph{medium}: moderate, ambiguous, or partial expression;
    \emph{high}: clear and strong expression of the construct.
Dimensions not expressed in the tweet are omitted rather than classified
as ``low'', and a single tweet may express multiple dimensions.

The qualitative labels are subsequently mapped onto a numerical scale:
$-1$ for \texttt{low}, $0.5$ for \texttt{medium}, and $1$ for
\texttt{high}; absence of the dimension is coded as $0$. Thus, for each
dimension, every tweet receives a value in $\{-1,0,0.5,1\}$.
The three-level qualitative classification is preferred to asking the
LLM to assign numerical scores directly, since LLMs are substantially
more reliable at categorical classification than at consistently
interpreting numerical scales
\citep{xu2024genius, zhang2024counting, fu2024counting}.
The classifier is based on the open-source Llama~3.2~3B model
\cite{llama32_3b}, fine-tuned using manually annotated tweets following
the procedure described in \cite{finetuning2026}. Fine-tuning specializes
the pretrained model to the classification task rather than training a
language model from scratch \citep{wei2022finetuned}. 

\clearpage

\subsection{Additional figures}

\subsubsection*{Heterogeneity in the within-region effect of left-behindness}

\begin{figure}[!h]
\centering
\includegraphics[width=\textwidth]{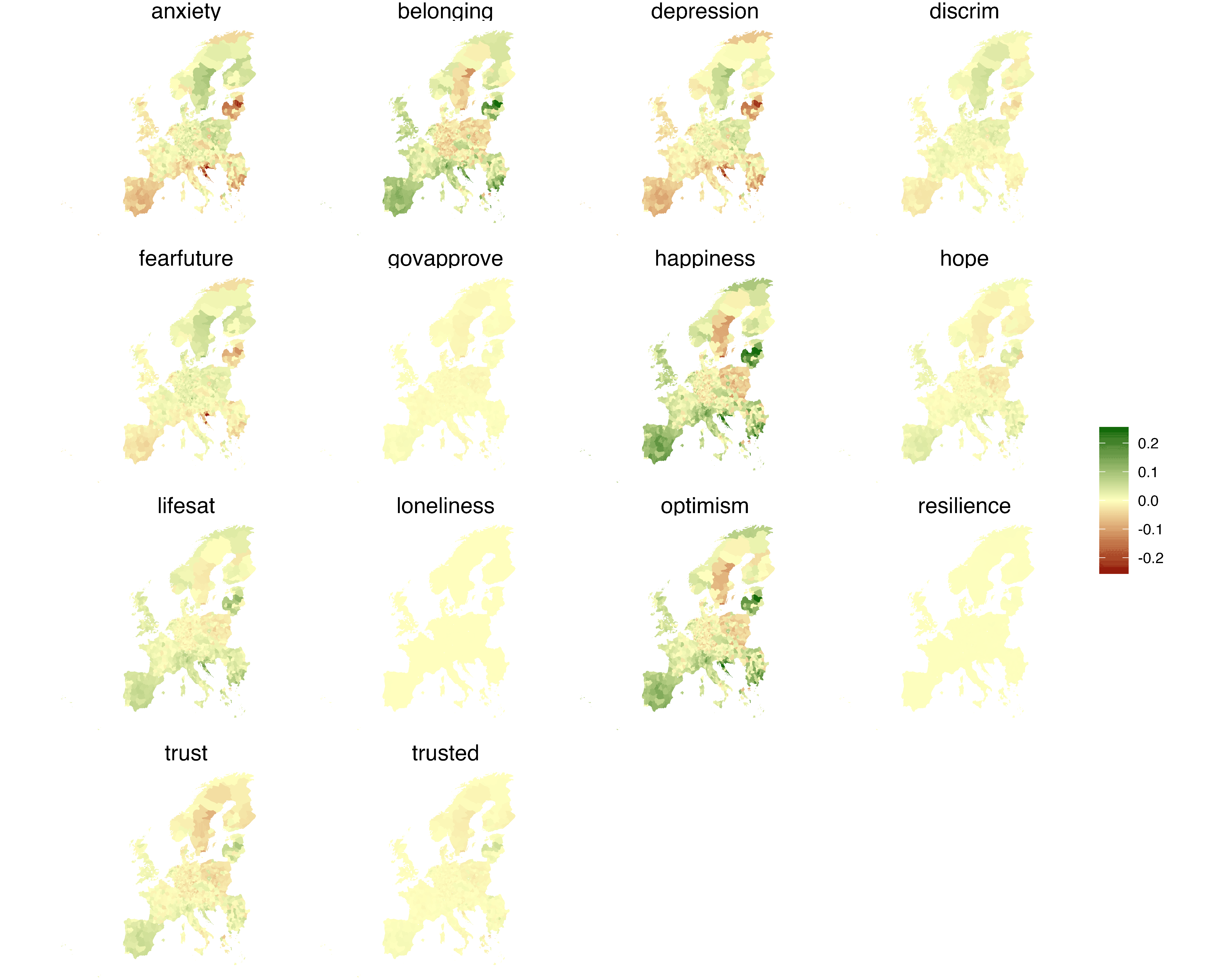}
\caption{Spatial heterogeneity of within-region effects of left-behindness score flourishing dimensions, EU+UK+EFTA NUTS-3. \textit{Note}: for visualization, color scale is clipped at the 80\% of maximum absolute value.}
\label{fig:map-randslp-dim}
\end{figure}

\section{Additional tables}
\subsubsection*{Full objective measures coefficient tables, all 18 outcomes}
\begin{table}[htbp]\centering\small
\caption{Model 1b (decomposed) within-between coefficients: happiness}
\label{tab:m1b-happiness}
\begin{tabular}{lccc}
\toprule
Regressor & Cross-national & Within-country & Within \\
 & (between) & (between) & (over time) \\
\midrule
GDP per capita relative to national mean & +0.041$^{**}$ & +0.021$^{**}$ & -0.002 \\
GDP per capita growth differential & +0.027$^{*}$ & +0.020$^{**}$ & +0.073$^{***}$ \\
Employment growth differential & -0.012 & +0.002 & +0.010 \\
Industrial employment share change & -0.067$^{***}$ & +0.000 & +0.022$^{***}$ \\
Net migration rate & -0.069$^{***}$ & -0.008 & -0.039$^{***}$ \\
Old-age dependency ratio & +0.027 & +0.017 & -0.408$^{***}$ \\
Population growth differential & +0.079$^{***}$ & +0.025 & -0.076$^{***}$ \\
At-risk-of-poverty rate & +0.063$^{***}$ & -0.004 & -0.026$^{***}$ \\
Accessibility index & -0.072$^{***}$ & -0.019 & +0.001 \\
Youth net migration rate & +0.032$^{*}$ & +0.022$^{*}$ & -- \\
\bottomrule
\end{tabular}
\end{table}

\begin{table}[htbp]\centering\small
\caption{Model 1b (decomposed) within-between coefficients: lifesat}
\label{tab:m1b-lifesat}
\begin{tabular}{lccc}
\toprule
Regressor & Cross-national & Within-country & Within \\
 & (between) & (between) & (over time) \\
\midrule
GDP per capita relative to national mean & +0.051$^{***}$ & +0.021$^{*}$ & +0.002 \\
GDP per capita growth differential & +0.007 & +0.007 & +0.051$^{***}$ \\
Employment growth differential & +0.002 & +0.010 & +0.009 \\
Industrial employment share change & -0.062$^{***}$ & +0.014 & +0.020$^{***}$ \\
Net migration rate & -0.110$^{***}$ & -0.005 & -0.036$^{***}$ \\
Old-age dependency ratio & +0.029 & +0.022 & -0.372$^{***}$ \\
Population growth differential & +0.112$^{***}$ & +0.026 & -0.076$^{***}$ \\
At-risk-of-poverty rate & +0.080$^{***}$ & -0.003 & -0.017$^{**}$ \\
Accessibility index & -0.102$^{***}$ & -0.019 & +0.001 \\
Youth net migration rate & +0.025 & +0.017 & -- \\
\bottomrule
\end{tabular}
\end{table}

\begin{table}[htbp]\centering\small
\caption{Model 1b (decomposed) within-between coefficients: anxiety}
\label{tab:m1b-anxiety}
\begin{tabular}{lccc}
\toprule
Regressor & Cross-national & Within-country & Within \\
 & (between) & (between) & (over time) \\
\midrule
GDP per capita relative to national mean & -0.017 & -0.008 & -0.006 \\
GDP per capita growth differential & -0.033$^{**}$ & -0.013 & -0.053$^{***}$ \\
Employment growth differential & +0.004 & -0.009 & -0.015 \\
Industrial employment share change & +0.056$^{***}$ & +0.000 & -0.030$^{***}$ \\
Net migration rate & +0.057$^{***}$ & +0.006 & +0.014$^{*}$ \\
Old-age dependency ratio & -0.045$^{**}$ & -0.032$^{*}$ & +0.483$^{***}$ \\
Population growth differential & -0.076$^{***}$ & -0.023 & +0.089$^{***}$ \\
At-risk-of-poverty rate & -0.087$^{***}$ & -0.002 & +0.019$^{**}$ \\
Accessibility index & -0.003 & +0.028 & +0.006 \\
Youth net migration rate & -0.033$^{*}$ & -0.010 & -- \\
\bottomrule
\end{tabular}
\end{table}

\begin{table}[htbp]\centering\small
\caption{Model 1b (decomposed) within-between coefficients: depression}
\label{tab:m1b-depression}
\begin{tabular}{lccc}
\toprule
Regressor & Cross-national & Within-country & Within \\
 & (between) & (between) & (over time) \\
\midrule
GDP per capita relative to national mean & -0.018 & -0.009 & -0.004 \\
GDP per capita growth differential & -0.032$^{**}$ & -0.016$^{*}$ & -0.059$^{***}$ \\
Employment growth differential & +0.001 & -0.014 & -0.009 \\
Industrial employment share change & +0.045$^{***}$ & -0.005 & -0.023$^{***}$ \\
Net migration rate & +0.046$^{**}$ & +0.001 & +0.024$^{***}$ \\
Old-age dependency ratio & -0.031$^{*}$ & -0.031$^{*}$ & +0.417$^{***}$ \\
Population growth differential & -0.059$^{**}$ & -0.014 & +0.080$^{***}$ \\
At-risk-of-poverty rate & -0.080$^{***}$ & +0.000 & +0.026$^{***}$ \\
Accessibility index & -0.006 & +0.021 & -0.001 \\
Youth net migration rate & -0.038$^{*}$ & -0.017 & -- \\
\bottomrule
\end{tabular}
\end{table}

\begin{table}[htbp]\centering\small
\caption{Model 1b (decomposed) within-between coefficients: belonging}
\label{tab:m1b-belonging}
\begin{tabular}{lccc}
\toprule
Regressor & Cross-national & Within-country & Within \\
 & (between) & (between) & (over time) \\
\midrule
GDP per capita relative to national mean & +0.051$^{***}$ & +0.023$^{**}$ & +0.001 \\
GDP per capita growth differential & +0.020 & +0.019$^{**}$ & +0.067$^{***}$ \\
Employment growth differential & +0.011 & +0.024$^{**}$ & +0.022$^{**}$ \\
Industrial employment share change & -0.051$^{***}$ & +0.007 & +0.043$^{***}$ \\
Net migration rate & -0.093$^{***}$ & +0.003 & -0.022$^{***}$ \\
Old-age dependency ratio & +0.048$^{***}$ & +0.028$^{*}$ & -0.465$^{***}$ \\
Population growth differential & +0.080$^{***}$ & -0.004 & -0.097$^{***}$ \\
At-risk-of-poverty rate & +0.062$^{***}$ & +0.005 & -0.023$^{***}$ \\
Accessibility index & -0.065$^{***}$ & +0.000 & +0.018$^{**}$ \\
Youth net migration rate & +0.043$^{***}$ & +0.048$^{***}$ & -- \\
\bottomrule
\end{tabular}
\end{table}

\begin{table}[htbp]\centering\small
\caption{Model 1b (decomposed) within-between coefficients: discrim}
\label{tab:m1b-discrim}
\begin{tabular}{lccc}
\toprule
Regressor & Cross-national & Within-country & Within \\
 & (between) & (between) & (over time) \\
\midrule
GDP per capita relative to national mean & -0.040$^{***}$ & -0.005 & -0.012 \\
GDP per capita growth differential & +0.013 & +0.008 & -0.018$^{**}$ \\
Employment growth differential & +0.003 & -0.013 & -0.040$^{***}$ \\
Industrial employment share change & +0.064$^{***}$ & -0.012 & -0.053$^{***}$ \\
Net migration rate & +0.038$^{**}$ & +0.005 & -0.035$^{***}$ \\
Old-age dependency ratio & -0.004 & -0.014 & +0.632$^{***}$ \\
Population growth differential & -0.026 & +0.009 & +0.105$^{***}$ \\
At-risk-of-poverty rate & -0.031$^{**}$ & -0.003 & +0.004 \\
Accessibility index & +0.102$^{***}$ & -0.004 & +0.017$^{**}$ \\
Youth net migration rate & +0.030$^{*}$ & -0.002 & -- \\
\bottomrule
\end{tabular}
\end{table}

\begin{table}[htbp]\centering\small
\caption{Model 1b (decomposed) within-between coefficients: govapprove}
\label{tab:m1b-govapprove}
\begin{tabular}{lccc}
\toprule
Regressor & Cross-national & Within-country & Within \\
 & (between) & (between) & (over time) \\
\midrule
GDP per capita relative to national mean & +0.043$^{***}$ & +0.000 & +0.022$^{**}$ \\
GDP per capita growth differential & -0.024$^{*}$ & -0.008 & -0.003 \\
Employment growth differential & -0.015 & +0.004 & +0.045$^{***}$ \\
Industrial employment share change & -0.075$^{***}$ & +0.002 & +0.067$^{***}$ \\
Net migration rate & -0.001 & +0.002 & +0.058$^{***}$ \\
Old-age dependency ratio & -0.041$^{**}$ & -0.005 & -0.617$^{***}$ \\
Population growth differential & -0.001 & -0.010 & -0.108$^{***}$ \\
At-risk-of-poverty rate & +0.012 & +0.011 & +0.001 \\
Accessibility index & -0.099$^{***}$ & +0.026 & -0.018$^{**}$ \\
Youth net migration rate & -0.071$^{***}$ & -0.005 & -- \\
\bottomrule
\end{tabular}
\end{table}

\begin{table}[htbp]\centering\small
\caption{Model 1b (decomposed) within-between coefficients: hope}
\label{tab:m1b-hope}
\begin{tabular}{lccc}
\toprule
Regressor & Cross-national & Within-country & Within \\
 & (between) & (between) & (over time) \\
\midrule
GDP per capita relative to national mean & +0.047$^{**}$ & +0.008 & +0.004 \\
GDP per capita growth differential & +0.029 & +0.015 & +0.047$^{***}$ \\
Employment growth differential & -0.014 & +0.010 & +0.017$^{*}$ \\
Industrial employment share change & -0.124$^{***}$ & +0.002 & +0.044$^{***}$ \\
Net migration rate & -0.015 & -0.030$^{*}$ & -0.009 \\
Old-age dependency ratio & -0.040 & +0.027 & -0.442$^{***}$ \\
Population growth differential & +0.039 & +0.044$^{**}$ & -0.081$^{***}$ \\
At-risk-of-poverty rate & +0.079$^{***}$ & -0.014 & -0.011 \\
Accessibility index & -0.095$^{***}$ & -0.021 & -0.017$^{**}$ \\
Youth net migration rate & -0.055$^{**}$ & -0.016 & -- \\
\bottomrule
\end{tabular}
\end{table}

\begin{table}[htbp]\centering\small
\caption{Model 1b (decomposed) within-between coefficients: loneliness}
\label{tab:m1b-loneliness}
\begin{tabular}{lccc}
\toprule
Regressor & Cross-national & Within-country & Within \\
 & (between) & (between) & (over time) \\
\midrule
GDP per capita relative to national mean & -0.010 & -0.020$^{*}$ & -0.011 \\
GDP per capita growth differential & -0.017 & -0.004 & -0.019$^{**}$ \\
Employment growth differential & -0.002 & -0.010 & +0.037$^{***}$ \\
Industrial employment share change & +0.020 & -0.020$^{*}$ & +0.020$^{**}$ \\
Net migration rate & +0.075$^{***}$ & -0.011 & +0.043$^{***}$ \\
Old-age dependency ratio & -0.050$^{**}$ & -0.008 & -0.012 \\
Population growth differential & -0.081$^{***}$ & +0.000 & -0.005 \\
At-risk-of-poverty rate & -0.105$^{***}$ & -0.003 & +0.035$^{***}$ \\
Accessibility index & +0.005 & +0.019 & +0.005 \\
Youth net migration rate & -0.094$^{***}$ & +0.012 & -- \\
\bottomrule
\end{tabular}
\end{table}

\begin{table}[htbp]\centering\small
\caption{Model 1b (decomposed) within-between coefficients: optimism}
\label{tab:m1b-optimism}
\begin{tabular}{lccc}
\toprule
Regressor & Cross-national & Within-country & Within \\
 & (between) & (between) & (over time) \\
\midrule
GDP per capita relative to national mean & +0.042$^{**}$ & +0.022$^{*}$ & -0.006 \\
GDP per capita growth differential & +0.034$^{*}$ & +0.025$^{**}$ & +0.078$^{***}$ \\
Employment growth differential & -0.010 & +0.003 & +0.011 \\
Industrial employment share change & -0.062$^{***}$ & +0.001 & +0.025$^{***}$ \\
Net migration rate & -0.065$^{***}$ & -0.009 & -0.041$^{***}$ \\
Old-age dependency ratio & +0.020 & +0.012 & -0.424$^{***}$ \\
Population growth differential & +0.082$^{***}$ & +0.028 & -0.079$^{***}$ \\
At-risk-of-poverty rate & +0.070$^{***}$ & -0.011 & -0.021$^{***}$ \\
Accessibility index & -0.068$^{***}$ & -0.030 & +0.001 \\
Youth net migration rate & +0.019 & +0.016 & -- \\
\bottomrule
\end{tabular}
\end{table}

\begin{table}[htbp]\centering\small
\caption{Model 1b (decomposed) within-between coefficients: resilience}
\label{tab:m1b-resilience}
\begin{tabular}{lccc}
\toprule
Regressor & Cross-national & Within-country & Within \\
 & (between) & (between) & (over time) \\
\midrule
GDP per capita relative to national mean & +0.038$^{*}$ & -0.001 & -0.009 \\
GDP per capita growth differential & +0.002 & +0.024$^{*}$ & +0.052$^{***}$ \\
Employment growth differential & -0.009 & +0.026 & +0.001 \\
Industrial employment share change & -0.072$^{***}$ & -0.008 & +0.026$^{***}$ \\
Net migration rate & +0.063$^{**}$ & -0.023 & -0.061$^{***}$ \\
Old-age dependency ratio & -0.153$^{***}$ & +0.026 & -0.021$^{*}$ \\
Population growth differential & -0.061$^{*}$ & +0.032 & -0.016 \\
At-risk-of-poverty rate & -0.022 & +0.004 & -0.032$^{***}$ \\
Accessibility index & -0.091$^{***}$ & +0.041 & -0.002 \\
Youth net migration rate & -0.085$^{***}$ & +0.010 & -- \\
\bottomrule
\end{tabular}
\end{table}

\begin{table}[htbp]\centering\small
\caption{Model 1b (decomposed) within-between coefficients: trust}
\label{tab:m1b-trust}
\begin{tabular}{lccc}
\toprule
Regressor & Cross-national & Within-country & Within \\
 & (between) & (between) & (over time) \\
\midrule
GDP per capita relative to national mean & +0.038$^{**}$ & +0.015 & +0.003 \\
GDP per capita growth differential & +0.034$^{**}$ & +0.019$^{*}$ & +0.066$^{***}$ \\
Employment growth differential & -0.027 & +0.005 & +0.016 \\
Industrial employment share change & -0.077$^{***}$ & -0.002 & +0.039$^{***}$ \\
Net migration rate & -0.064$^{***}$ & -0.013 & +0.002 \\
Old-age dependency ratio & -0.010 & +0.005 & -0.538$^{***}$ \\
Population growth differential & +0.060$^{**}$ & +0.015 & -0.079$^{***}$ \\
At-risk-of-poverty rate & +0.054$^{***}$ & -0.008 & -0.013$^{*}$ \\
Accessibility index & -0.086$^{***}$ & -0.019 & -0.005 \\
Youth net migration rate & -0.015 & +0.013 & -- \\
\bottomrule
\end{tabular}
\end{table}

\begin{table}[htbp]\centering\small
\caption{Model 1b (decomposed) within-between coefficients: trusted}
\label{tab:m1b-trusted}
\begin{tabular}{lccc}
\toprule
Regressor & Cross-national & Within-country & Within \\
 & (between) & (between) & (over time) \\
\midrule
GDP per capita relative to national mean & +0.036$^{**}$ & +0.014 & +0.016$^{*}$ \\
GDP per capita growth differential & +0.045$^{***}$ & +0.013 & +0.028$^{***}$ \\
Employment growth differential & -0.038$^{*}$ & -0.007 & +0.009 \\
Industrial employment share change & -0.061$^{***}$ & +0.013 & +0.052$^{***}$ \\
Net migration rate & -0.038$^{*}$ & -0.005 & +0.022$^{**}$ \\
Old-age dependency ratio & -0.019 & +0.003 & -0.474$^{***}$ \\
Population growth differential & +0.045$^{*}$ & +0.000 & -0.081$^{***}$ \\
At-risk-of-poverty rate & +0.092$^{***}$ & +0.001 & -0.039$^{***}$ \\
Accessibility index & +0.002 & +0.010 & -0.006 \\
Youth net migration rate & -0.003 & +0.000 & -- \\
\bottomrule
\end{tabular}
\end{table}

\begin{table}[htbp]\centering\small
\caption{Model 1b (decomposed) within-between coefficients: fearfuture}
\label{tab:m1b-fearfuture}
\begin{tabular}{lccc}
\toprule
Regressor & Cross-national & Within-country & Within \\
 & (between) & (between) & (over time) \\
\midrule
GDP per capita relative to national mean & -0.027$^{*}$ & -0.010 & -0.014$^{*}$ \\
GDP per capita growth differential & -0.023 & -0.019$^{*}$ & -0.029$^{***}$ \\
Employment growth differential & +0.016 & -0.007 & -0.024$^{**}$ \\
Industrial employment share change & +0.063$^{***}$ & +0.011 & -0.047$^{***}$ \\
Net migration rate & +0.062$^{***}$ & +0.008 & -0.014$^{*}$ \\
Old-age dependency ratio & -0.004 & -0.019 & +0.475$^{***}$ \\
Population growth differential & -0.081$^{***}$ & -0.025 & +0.093$^{***}$ \\
At-risk-of-poverty rate & -0.089$^{***}$ & +0.008 & +0.015$^{*}$ \\
Accessibility index & -0.011 & +0.024 & +0.002 \\
Youth net migration rate & +0.011 & -0.016 & -- \\
\bottomrule
\end{tabular}
\end{table}

\begin{table}[htbp]\centering\small
\caption{Model 1b (decomposed) within-between coefficients: composite}
\label{tab:m1b-composite}
\begin{tabular}{lccc}
\toprule
Regressor & Cross-national & Within-country & Within \\
 & (between) & (between) & (over time) \\
\midrule
GDP per capita relative to national mean & +0.043$^{***}$ & +0.015$^{*}$ & +0.007 \\
GDP per capita growth differential & +0.023 & +0.015$^{*}$ & +0.055$^{***}$ \\
Employment growth differential & -0.012 & +0.011 & +0.017$^{*}$ \\
Industrial employment share change & -0.078$^{***}$ & +0.005 & +0.041$^{***}$ \\
Net migration rate & -0.059$^{***}$ & -0.008 & -0.013$^{*}$ \\
Old-age dependency ratio & +0.001 & +0.021 & -0.500$^{***}$ \\
Population growth differential & +0.066$^{**}$ & +0.018 & -0.092$^{***}$ \\
At-risk-of-poverty rate & +0.077$^{***}$ & -0.002 & -0.024$^{***}$ \\
Accessibility index & -0.066$^{***}$ & -0.011 & -0.005 \\
Youth net migration rate & +0.002 & +0.012 & -- \\
\bottomrule
\end{tabular}
\end{table}

\begin{table}[htbp]\centering\small
\caption{Model 1b (decomposed) within-between coefficients: composite\_cohesion}
\label{tab:m1b-compositecohesion}
\begin{tabular}{lccc}
\toprule
Regressor & Cross-national & Within-country & Within \\
 & (between) & (between) & (over time) \\
\midrule
GDP per capita relative to national mean & +0.048$^{***}$ & +0.015$^{*}$ & +0.012 \\
GDP per capita growth differential & +0.021$^{*}$ & +0.012 & +0.045$^{***}$ \\
Employment growth differential & -0.020 & +0.007 & +0.026$^{***}$ \\
Industrial employment share change & -0.075$^{***}$ & +0.006 & +0.057$^{***}$ \\
Net migration rate & -0.056$^{***}$ & -0.004 & +0.017$^{**}$ \\
Old-age dependency ratio & -0.006 & +0.009 & -0.594$^{***}$ \\
Population growth differential & +0.052$^{**}$ & +0.000 & -0.104$^{***}$ \\
At-risk-of-poverty rate & +0.062$^{***}$ & +0.003 & -0.021$^{***}$ \\
Accessibility index & -0.071$^{***}$ & +0.005 & -0.003 \\
Youth net migration rate & -0.013 & +0.016 & -- \\
\bottomrule
\end{tabular}
\end{table}

\begin{table}[htbp]\centering\small
\caption{Model 1b (decomposed) within-between coefficients: composite\_wellbeing}
\label{tab:m1b-compositewellbeing}
\begin{tabular}{lccc}
\toprule
Regressor & Cross-national & Within-country & Within \\
 & (between) & (between) & (over time) \\
\midrule
GDP per capita relative to national mean & +0.051$^{***}$ & +0.017$^{*}$ & -0.002 \\
GDP per capita growth differential & +0.023 & +0.021$^{**}$ & +0.071$^{***}$ \\
Employment growth differential & -0.010 & +0.012 & +0.011 \\
Industrial employment share change & -0.091$^{***}$ & +0.002 & +0.032$^{***}$ \\
Net migration rate & -0.048$^{**}$ & -0.018 & -0.043$^{***}$ \\
Old-age dependency ratio & -0.025 & +0.024 & -0.398$^{***}$ \\
Population growth differential & +0.061$^{*}$ & +0.036$^{*}$ & -0.078$^{***}$ \\
At-risk-of-poverty rate & +0.065$^{***}$ & -0.007 & -0.025$^{***}$ \\
Accessibility index & -0.101$^{***}$ & -0.012 & -0.004 \\
Youth net migration rate & -0.014 & +0.011 & -- \\
\bottomrule
\end{tabular}
\end{table}

\begin{table}[htbp]\centering\small
\caption{Model 1b (decomposed) within-between coefficients: composite\_illbeing}
\label{tab:m1b-compositeillbeing}
\begin{tabular}{lccc}
\toprule
Regressor & Cross-national & Within-country & Within \\
 & (between) & (between) & (over time) \\
\midrule
GDP per capita relative to national mean & -0.026$^{*}$ & -0.012 & -0.011 \\
GDP per capita growth differential & -0.022 & -0.010 & -0.042$^{***}$ \\
Employment growth differential & +0.005 & -0.013 & -0.012 \\
Industrial employment share change & +0.058$^{***}$ & -0.006 & -0.031$^{***}$ \\
Net migration rate & +0.065$^{***}$ & +0.002 & +0.008 \\
Old-age dependency ratio & -0.031$^{*}$ & -0.024 & +0.464$^{***}$ \\
Population growth differential & -0.075$^{***}$ & -0.012 & +0.084$^{***}$ \\
At-risk-of-poverty rate & -0.091$^{***}$ & +0.000 & +0.023$^{***}$ \\
Accessibility index & +0.020 & +0.021 & +0.007 \\
Youth net migration rate & -0.029 & -0.008 & -- \\
\bottomrule
\end{tabular}
\end{table}

\end{document}